# The Dark Energy Survey Data Release 2

T. M. C. Abbott[1], M. Adamów[2], M. Aguena[3,4], S. Allam[5], A. Amon[6], J. Annis[5], S. Avila[7], D. Bacon[8], M. Banerji[9,10], K. Bechtol[11], M. R. Becker[12], G. M. Bernstein[13], E. Bertin[14,15], S. Bhargava[16], S. L. Bridle[17], D. Brooks[18], D. L. Burke[6,19], A. Carnero Rosell[4,20,21], M. Carrasco Kind[2,22], J. Carretero[23], F. J. Castander[24,25], R. Cawthon[11], C. Chang[26,27], A. Choi[28], C. Conselice[17,29], M. Costanzi[30,31], M. Crocce[24,25], L. N. da Costa[4,32], T. M. Davis[33], J. De Vicente[34], J. DeRose[35,36], S. Desai[37], H. T. Diehl[5], J. P. Dietrich[38], A. Drlica-Wagner[5,26,27], K. Eckert[13], J. Elvin-Poole[28,39], S. Everett[36], A. E. Evrard[40,41], I. Ferrero[42], A. Ferté[43], B. Flaugher[5], P. Fosalba[24,25], D. Friedel[2], J. Frieman[5,27], J. García-Bellido[7], E. Gaztanaga[24,25], L. Gelman[2], D. W. Gerdes[40,41], T. Giannantonio[9], M. S. S. Gill[19], D. Gruen[6,19,44], R. A. Gruendl[2,22], J. Gschwend[4,32], G. Gutierrez[5], W. G. Hartley[45], S. R. Hinton[33], D. L. Hollowood[36], K. Honscheid[28,39], D. Huterer[41], D. J. James[46], T. Jeltema[36], M. D. Johnson[2], S. Kent[5,27], R. Kron[5,26], K. Kuehn[47,48], N. Kuropatkin[5], O. Lahav[18], T. S. Li[49,50], C. Lidman[51,52], H. Lin[5], N. MacCrann[53], M. A. G. Maia[4,32], T. A. Manning[2], J. D. Maloney[2], M. March[13], J. L. Marshall[54], P. Martini[28,55,56], P. Melchior[49], F. Menanteau[2,22], R. Miquel[23,57], R. Morgan[11], J. Myles[6,19,44], E. Neilsen[5], R. L. C. Ogando[4,32], A. Palmese[5,27], F. Paz-Chinchón[2,9], D. Petravick[2], A. Pieres[4,32], A. A. Plazas[49], C. Pond[2], M. Rodriguez-Monroy[34], A. K. Romer[16], A. Roodman[6,19], E. S. Rykoff[6,19], M. Sako[13], E. Sanchez[34], B. Santiago[4,58], V. Scarpine[5], S. Serrano[24,25], I. Sevilla-Noarbe[34], J. Allyn Smith[59], M. Smith[60], M. Soares-Santos[41], E. Suchyta[61], M. E. C. Swanson[2], G. Tarle[41], D. Thomas[8], C. To[6,19,44], P. E. Tremblay[62], M. A. Troxel[63], D. L. Tucker[5], D. J. Turner[16], T. N. Varga[64,65], A. R. Walker[1], R. H. Wechsler[6,19,44], J. Weller[64,65], W. Wester[5], R. D. Wilkinson[16], B. Yanny[5], Y. Zhang[5], R. Nikutta[66], M. Fitzpatrick[66], A. Jacques[66], A. Scott[66], K. Olsen[66], L. Huang[66], D. Herrera[66], S. Juneau[66], D. Nidever[66,67], B. A. Weaver[66], C. Adean[4], V. Correia[4], M. de Freitas[4], F. N. Freitas[4], C. Singulani[4], and G. Vila-Verde[4]
(Linea Science Server)

[1] Cerro Tololo Inter-American Observatory, NSF's National Optical-Infrared Astronomy Research Laboratory, Casilla 603, La Serena, Chile
[2] Center for Astrophysical Surveys, National Center for Supercomputing Applications, 1205 West Clark Street, Urbana, IL 61801, USA; mcarras2@illinois.edu
[3] Departamento de Física Matemática, Instituto de Física, Universidade de São Paulo, CP 66318, São Paulo, SP, 05314-970, Brazil
[4] Laboratório Interinstitucional de e-Astronomia - LIneA, Rua Gal. José Cristino 77, Rio de Janeiro, RJ - 20921-400, Brazil; aurelio.crosell@gmail.com
[5] Fermi National Accelerator Laboratory, P.O. Box 500, Batavia, IL 60510, USA
[6] Kavli Institute for Particle Astrophysics & Cosmology, P.O. Box 2450, Stanford University, Stanford, CA 94305, USA
[7] Instituto de Fisica Teorica UAM/CSIC, Universidad Autonoma de Madrid, E-28049 Madrid, Spain
[8] Institute of Cosmology and Gravitation, University of Portsmouth, Portsmouth, PO1 3FX, UK
[9] Institute of Astronomy, University of Cambridge, Madingley Road, Cambridge, CB3 0HA, UK
[10] Kavli Institute for Cosmology, University of Cambridge, Madingley Road, Cambridge, CB3 0HA, UK
[11] Physics Department, 2320 Chamberlin Hall, University of Wisconsin-Madison, 1150 University Avenue, Madison, WI 53706-1390, USA; kbechtol@wisc.edu
[12] Argonne National Laboratory, 9700 South Cass Avenue, Lemont, IL 60439, USA
[13] Department of Physics and Astronomy, University of Pennsylvania, Philadelphia, PA 19104, USA
[14] CNRS, UMR 7095, Institut d'Astrophysique de Paris, F-75014, Paris, France
[15] Sorbonne Universités, UPMC Univ Paris 06, UMR 7095, Institut d'Astrophysique de Paris, F-75014, Paris, France
[16] Department of Physics and Astronomy, Pevensey Building, University of Sussex, Brighton, BN1 9QH, UK
[17] Jodrell Bank Center for Astrophysics, School of Physics and Astronomy, University of Manchester, Oxford Road, Manchester, M13 9PL, UK
[18] Department of Physics & Astronomy, University College London, Gower Street, London, WC1E 6BT, UK
[19] SLAC National Accelerator Laboratory, Menlo Park, CA 94025, USA
[20] Instituto de Astrofísica de Canarias, E-38205 La Laguna, Tenerife, Spain
[21] Universidad de La Laguna, Dpto. Astrofísica, E-38206 La Laguna, Tenerife, Spain
[22] Department of Astronomy, University of Illinois at Urbana-Champaign, 1002 W. Green Street, Urbana, IL 61801, USA
[23] Institut de Física d'Altes Energies (IFAE), The Barcelona Institute of Science and Technology, Campus UAB, E-08193 Bellaterra (Barcelona) Spain
[24] Institut d'Estudis Espacials de Catalunya (IEEC), E-08034 Barcelona, Spain
[25] Institute of Space Sciences (ICE, CSIC), Campus UAB, Carrer de Can Magrans, s/n, E-08193 Barcelona, Spain
[26] Department of Astronomy and Astrophysics, University of Chicago, Chicago, IL 60637, USA
[27] Kavli Institute for Cosmological Physics, University of Chicago, Chicago, IL 60637, USA
[28] Center for Cosmology and Astro-Particle Physics, The Ohio State University, Columbus, OH 43210, USA
[29] University of Nottingham, School of Physics and Astronomy, Nottingham, NG7 2RD, UK
[30] INAF-Osservatorio Astronomico di Trieste, via G.B. Tiepolo 11, I-34143 Trieste, Italy
[31] Institute for Fundamental Physics of the Universe, Via Beirut 2, I-34014 Trieste, Italy
[32] Observatório Nacional, Rua Gal. José Cristino 77, Rio de Janeiro, RJ - 20921-400, Brazil
[33] School of Mathematics and Physics, University of Queensland, Brisbane, QLD 4072, Australia
[34] Centro de Investigaciones Energéticas, Medioambientales y Tecnológicas (CIEMAT), Madrid, Spain
[35] Department of Astronomy, University of California, Berkeley, 501 Campbell Hall, Berkeley, CA 94720, USA
[36] Santa Cruz Institute for Particle Physics, Santa Cruz, CA 95064, USA
[37] Department of Physics, IIT Hyderabad, Kandi, Telangana 502285, India
[38] Faculty of Physics, Ludwig-Maximilians-Universität, Scheinerstr. 1, D-81679 Munich, Germany
[39] Department of Physics, The Ohio State University, Columbus, OH 43210, USA
[40] Department of Astronomy, University of Michigan, Ann Arbor, MI 48109, USA
[41] Department of Physics, University of Michigan, Ann Arbor, MI 48109, USA
[42] Institute of Theoretical Astrophysics, University of Oslo, P.O. Box 1029 Blindern, NO-0315 Oslo, Norway
[43] Jet Propulsion Laboratory, California Institute of Technology, 4800 Oak Grove Drive, Pasadena, CA 91109, USA
[44] Department of Physics, Stanford University, 382 Via Pueblo Mall, Stanford, CA 94305, USA

[45] Department of Astronomy, University of Geneva, ch. d'Ecogia 16, CH-1290 Versoix, Switzerland
[46] Center for Astrophysics | Harvard & Smithsonian, 60 Garden Street, Cambridge, MA 02138, USA
[47] Australian Astronomical Optics, Macquarie University, North Ryde, NSW 2113, Australia
[48] Lowell Observatory, 1400 Mars Hill Road, Flagstaff, AZ 86001, USA
[49] Department of Astrophysical Sciences, Princeton University, Peyton Hall, Princeton, NJ 08544, USA
[50] Observatories of the Carnegie Institution for Science, 813 Santa Barbara Street, Pasadena, CA 91101, USA
[51] Centre for Gravitational Astrophysics, College of Science, The Australian National University, ACT 2601, Australia
[52] The Research School of Astronomy and Astrophysics, The Australian National University, ACT 2601, Australia
[53] Department of Applied Mathematics and Theoretical Physics, University of Cambridge, Cambridge, CB3 0WA, UK
[54] George P. and Cynthia Woods Mitchell Institute for Fundamental Physics and Astronomy, and Department of Physics and Astronomy, Texas A&M University, College Station, TX 77843, USA
[55] Department of Astronomy, The Ohio State University, Columbus, OH 43210, USA
[56] Radcliffe Institute for Advanced Study, Harvard University, Cambridge, MA 02138, USA
[57] Institució Catalana de Recerca i Estudis Avançats, E-08010 Barcelona, Spain
[58] Instituto de Física, UFRGS, Caixa Postal 15051, Porto Alegre, RS - 91501-970, Brazil
[59] Austin Peay State University, Department of Physics, Engineering and Astronomy, P.O. Box 4608, Clarksville, TN 37044, USA
[60] School of Physics and Astronomy, University of Southampton, Southampton, SO17 1BJ, UK
[61] Computer Science and Mathematics Division, Oak Ridge National Laboratory, Oak Ridge, TN 37831, USA
[62] Department of Physics, University of Warwick, Coventry, CV4 7AL, UK
[63] Department of Physics, Duke University Durham, NC 27708, USA
[64] Max Planck Institute for Extraterrestrial Physics, Giessenbachstrasse, D-85748 Garching, Germany
[65] Universitäts-Sternwarte, Fakultät für Physik, Ludwig-Maximilians Universität München, Scheinerstr. 1, D-81679 München, Germany
[66] NSF's National Optical-Infrared Astronomy Research Laboratory (NOIRLab), 950 N Cherry Avenue, Tucson, AZ 85719, USA
[67] Department of Physics, Montana State University, P.O. Box 173840, Bozeman, MT 59717-3840, USA

## Abstract

We present the second public data release of the Dark Energy Survey, DES DR2, based on optical/near-infrared imaging by the Dark Energy Camera mounted on the 4 m Blanco telescope at Cerro Tololo Inter-American Observatory in Chile. DES DR2 consists of reduced single-epoch and coadded images, a source catalog derived from coadded images, and associated data products assembled from 6 yr of DES science operations. This release includes data from the DES wide-area survey covering $\sim$5000 deg$^2$ of the southern Galactic cap in five broad photometric bands, $grizY$. DES DR2 has a median delivered point-spread function FWHM of $g = 1.11''$, $r = 0.95''$, $i = 0.88''$, $z = 0.83''$, and $Y = 0.''90$, photometric uniformity with a standard deviation of $< 3$ mmag with respect to Gaia DR2 $G$ band, a photometric accuracy of $\sim$11 mmag, and a median internal astrometric precision of $\sim$27 mas. The median coadded catalog depth for a $1.''95$ diameter aperture at signal-to-noise ratio = 10 is $g = 24.7$, $r = 24.4$, $i = 23.8$, $z = 23.1$, and $Y = 21.7$ mag. DES DR2 includes $\sim$691 million distinct astronomical objects detected in 10,169 coadded image tiles of size 0.534 deg$^2$ produced from 76,217 single-epoch images. After a basic quality selection, benchmark galaxy and stellar samples contain 543 million and 145 million objects, respectively. These data are accessible through several interfaces, including interactive image visualization tools, web-based query clients, image cutout servers, and Jupyter notebooks. DES DR2 constitutes the largest photometric data set to date at the achieved depth and photometric precision.

*Key words:* Dark energy – Cosmology – Extragalactic astronomy – Surveys – Redshift surveys – Optical astronomy – Near infrared astronomy

## 1. Introduction

The Dark Energy Survey (DES) is a ground-based, wide-area visible and near-infrared imaging survey covering $\sim 5000 \text{ deg}^2$ of the southern high Galactic latitude sky. DES was designed to improve our understanding of cosmic acceleration and the nature of dark energy using four complementary measurements: weak gravitational lensing, galaxy cluster counts, the large-scale clustering of galaxies (including baryon acoustic oscillations), and the distances to Type Ia supernovae (DES Collaboration 2005, 2019a). To achieve these goals, DES has conducted two distinct surveys: a $\sim$5000 deg$^2$ wide-area survey in the $grizY$ bands and a $\sim$27 deg$^2$ deep supernova survey observed in the $griz$ bands with a $\sim$7 day cadence (Kessler et al. 2015; Diehl et al. 2019). The design and unprecedented sensitivity of DES facilitate the study of many other astrophysical science cases beyond dark energy (e.g., DES Collaboration 2016).

To achieve the DES science goals, the DES Collaboration designed, built, and utilized the Dark Energy Camera (DECam; Honscheid et al. 2008; Flaugher et al. 2015), a 570 megapixel camera with a 3 deg$^2$ field of view installed at the prime focus of the 4 m Blanco telescope at Cerro Tololo Inter-American Observatory (CTIO) in northern Chile. During the first five years of survey observations (Y1–Y5), DES was allocated $\sim$105 equivalent full nights per year (August through mid-February), which were observed as a combination of full and half nights. The sixth year of observations (Y6) comprised 52 equivalent nights (Diehl et al. 2019), which were mostly observed as half nights. Each DES exposure was delivered from CTIO to the National Center for Supercomputing Applications (NCSA) at the University of Illinois at Urbana-Champaign for initial processing, generally within minutes of being observed. At NCSA, the DES Data Management system (DESDM; Sevilla et al. 2011; Desai et al. 2012; Mohr et al. 2012; Morganson et al. 2018) generates a variety of scientific products including initial rapid reductions ("First Cut"), final single-epoch reductions ("Final Cut"), and coadd images with associated source catalogs of suitable quality to perform precise cosmological measurements (e.g., DES Collaboration 2018a, 2019a, 2019b). Raw DES exposures are made publicly available one year after acquisition through the NSF's National Optical-Infrared Astronomy Research Laboratory (NOIRLab) Astro Data Archive.[68]

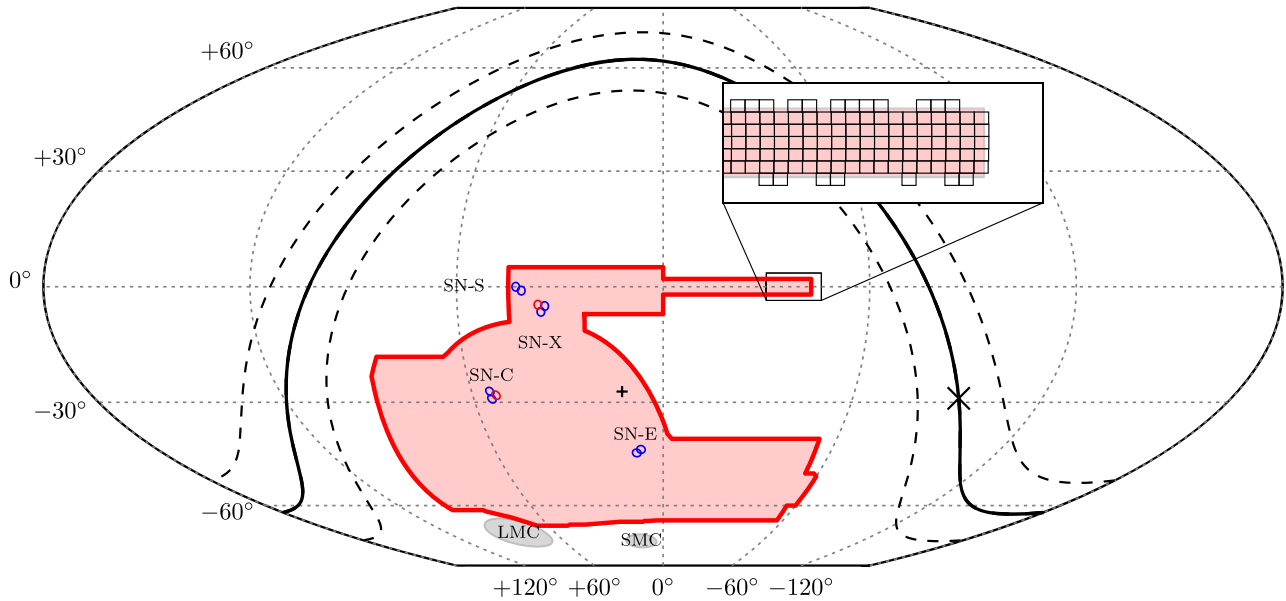

**Figure 1.** DES footprint in equatorial coordinates. The ∼5000 deg$^2$ wide-area survey footprint is shown in red. The eight shallow supernova fields are shown as blue circles, and the two deep supernova fields are shown as red circles. The Milky Way plane is shown as a solid line, with dashed lines at $b = \pm 10°$. The Galactic center ("×") and south Galactic pole ("+") are also marked. The Large and Small Magellanic Clouds are indicated in gray at the bottom of the footprint. The inset panel shows an overlay of tiles (coadded images, shown in black) on top of the western extreme of the SDSS Stripe 82 area. This and the other sky-map plots included in this work use the equal-area McBryde-Thomas flat-polar quartic projection.

The first major release of processed DES data, DES DR1, encompassed data products derived from the wide-area and the supernova surveys taken in the first three years of science operations (2013 August–2016 February) and was made public in 2018 January (DES Collaboration 2018b). In addition to the planned public data releases, the DES Collaboration prepares incremental releases with value-added products and detailed characterizations of survey performance that are designed to support cosmological analyses (e.g., the "Gold" catalogs; Drlica-Wagner et al. 2018; Sevilla-Noarbe et al. 2021). The value-added products associated with the DES Science Verification (SV) period (2012 November 1 through 2013 February 22) were released in 2016 January,[69] those associated with selected DES Y1 publications in September 2018[70], and those supporting the DES Y3 cosmological analyses are expected in 2021. Future releases are expected to support the Y6 scientific publications.

Here we present the content, validation, and data access services for the second public DES data release, DES DR2. DES DR2 comprises coadded images of size 0.534 deg$^2$ and source catalogs, as well as calibrated single-epoch CCD images, from the processing of all six years of DES wide-area survey observations and all five years of DES supernova survey observations. DES DR2 contains roughly 18 TB of coadd images and 3 TB of tables. All DES DR1 data have been reprocessed as input into DES DR2.

Access to DES DR2 data is provided via web interfaces and auxiliary tools, which are made possible through the partnership between NCSA, the Laboratório Interinstitucional de e-Astronomia (LIneA), and NOIRLab. The primary landing page for DES DR2 data products is https://des.ncsa.illinois.edu/releases/dr2.

In Section 2 we briefly summarize DES operations and data taking. In Section 3, we summarize the DES data processing and describe pipeline changes from DES DR1 to DES DR2. Basic validation of DES DR2 is presented in Section 4. Section 5 describes the products, data access frameworks, and tools made available for DES DR2. A summary of the release is given in Section 6. The Appendices present photometric transformation equations between DES DR2 and other large-area surveys, details on the absolute photometric calibration, data processing parameters, and a detailed description of the data products.

All magnitudes quoted in the text are in the AB system (Oke 1974), and all astronomical coordinates are provided in the Gaia-CRF2 reference frame (Gaia Collaboration 2018). All quoted uncertainties are $1\sigma$ (roughly 68% confidence interval) unless indicated otherwise.

## 2. DES Operations

DES was scheduled to observe for 760 distinct full or half nights between 2013 August 15 and 2019 January 9 (Diehl et al. 2019). DES observations are divided between a wide-area imaging program and a time-domain program (Figure 1). The DES DR2 coadd images and catalogs are derived from a subset of the wide-area component of DES and include exposures collected during 681 distinct nights.

The wide-area footprint was designed to significantly overlap with the South Pole Telescope survey (Carlstrom et al. 2011) and SDSS Stripe 82 (Abazajian et al. 2009), and it includes a connection region to enhance overall calibration.

---
[68] https://astroarchive.noirlab.edu/
[69] https://des.ncsa.illinois.edu/releases/sva1
[70] https://des.ncsa.illinois.edu/releases/y1a1

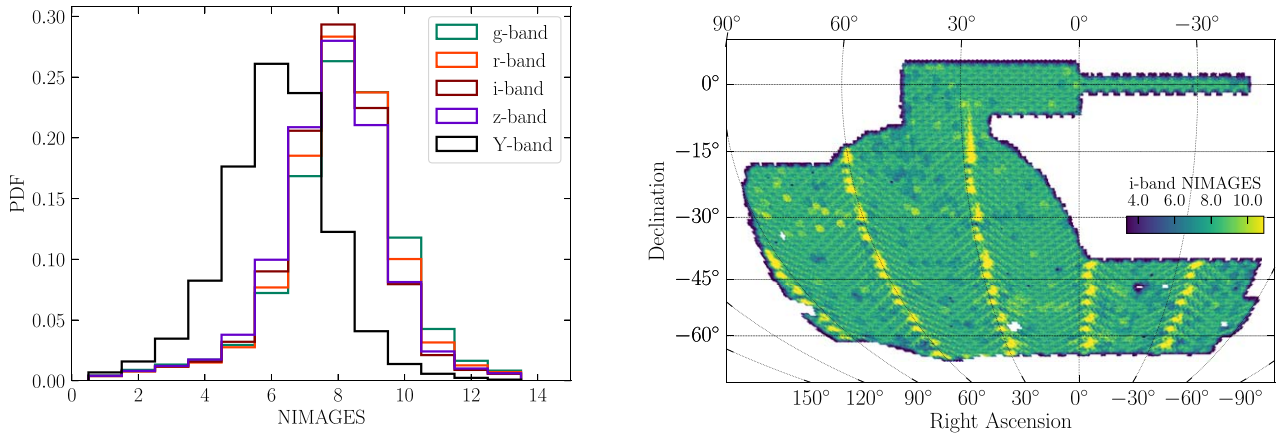

**Figure 2.** (Left) Normalized distribution of overlapping *grizY* single-epoch images in the DES DR2 wide-area survey footprint. Image overlap is averaged over nside = 4096 `HEALPix` pixels. Most regions of the footprint are covered with 7–10 images. (Right) DES DR2 footprint showing the number of overlapping *i*-band images. The regions containing more images spaced every ∼30° in R.A. are a result of the DES tiling scheme (Neilsen et al. 2019).

Given the cosmological goals of the survey, DES avoids the Galactic plane to minimize stellar foregrounds and extinction from interstellar dust. Observations were scheduled using an automated tool intended to maximize uniformity and depth over the full DES program (Neilsen et al. 2019).

DES collected ∼300 science exposures during a full night of observing, with variations depending on the season and survey strategy. These raw data were transferred to NOIRLab on a nightly basis for archiving and public distribution (Fitzpatrick 2010; Honscheid et al. 2012). The wide-area survey was observed with 90 s exposures in *griz*. The *Y*-band observations used a 45 s exposure time for Y1–Y3 and a 90 s exposure time in Y4–Y6.[71] The median overhead between DES wide-area survey exposures is 29 s, which includes readout, filter change, and slew. Each position in the DES DR2 footprint was typically observed in 7–10 overlapping DECam exposures in the *griz* bands (Figure 2). A more complete description of the DES survey strategy can be found in Neilsen et al. (2019).

DES DR2 includes a total of 96,263 DECam exposures (each comprising 62 science CCDs), including 83,706 exposures acquired as part of the wide-area survey and 12,557 exposures acquired as part of the supernova survey and other programs. A total of 76,217 exposures from the wide-area survey pass the baseline survey quality thresholds based on effective exposure time and image quality and are included in coadd processing by DESDM (Morganson et al. 2018). In this work, all quoted data quality characteristics (e.g., Table 1) refer to the subset of exposures included in the coadd unless stated otherwise. The DES DR2 exposures have a median airmass of 1.2, with >99% of exposures having airmass <1.4. The median delivered point-spread function (PSF) FWHM in units of arcsecond is $g = 1.11$, $r = 0.95$, $i = 0.88$, $z = 0.83$, and $Y = 0.90$. Zhang et al. (2019) investigated the extended wings of the PSF in the *g* and *r* bands and found that beyond the central $1\rlap{.}''1$, it can be well modeled by a three-component broken power law out to large angular separations (up to $\theta < 450''$).

For individual bands, the areas covered by at least one single-epoch image in that band are $g = 5055$, $r = 4988$, $i = 5023$, $z = 5031$, and $Y = 5080$ deg$^2$. When requiring at least one single-epoch image in each of the *grizY* bands, the DES DR2 footprint covers 4913 deg$^2$. The quoted areas exclude masked regions around bright stars and other imaging artifacts (Section 4.4).

### 3. Data Processing

The DES image reduction, detrending, and processing were performed by the DESDM system at NCSA (Morganson et al. 2018). Here we briefly summarize the processing pipeline and describe changes that were implemented in DES DR2 relative to previous data releases (i.e., DES Collaboration 2018b).

The DES Final Cut single-epoch processing pipeline removes instrumental signatures from individual exposures to produce reduced, science-ready, single-epoch CCD images. The Final Cut pipeline performs flat-fielding, overscan removal, crosstalk correction, nonlinearity correction, bias subtraction, gain correction, correction for the brighter-fatter effect, bad-pixel masking, astrometric matching, flagging of saturated pixels and bleed trails, principal-components background subtraction, secondary flat-field correction, and the masking of cosmic rays, artificial satellites, and other imaging artifacts.

The resulting processed images from this pipeline are released through the NOIRLab Astro Data Archive. Those images are provided as FITS-formatted files and contain extensions for the science data (`SCI`), a mask of bad pixels (`MSK`), and an inverse-variance weight (`WGT`). As in DR1, the weight images do not include flagged defects. A summary of the flags is provided in Table 9 of Morganson et al. (2018).

The Final Cut pipeline also performs PSF model fitting with `PSFEx` (Bertin 2011) and source detection and measurement with `SourceExtractor` (Bertin & Arnouts 1996). Photometric zero points (ZP) for the reduced single-epoch images are derived with the Forward Global Calibration Method (FGCM; Burke et al. 2018). The values of the single-epoch image pixels, *I*(pix), are provided in units of total corrected photoelectrons and can be converted to flux density in units of nJy as

$$F = 3.63 \times I(\text{pix}) \times 10^{((30-ZP)/2.5)} \text{ [nJy]}. \quad (1)$$

The typical depth of the Final Cut images estimated from the PSF magnitude at a signal-to-noise ratio (S/N) = 10 is $g = 23.57$, $r = 23.34$, $i = 22.78$, $z = 22.10$, and $Y = 20.69$ (Morganson et al. 2018). At the other extreme, a DECam

---
[71] The change in the *Y*-band exposure time was made to increase survey efficiency while retaining the same cumulative exposure time.

**Table 1**
DES DR2 Key Numbers and Data Quality Summary

| Parameter | Band | | | | | Reference |
|---|---|---|---|---|---|---|
| | g | r | i | z | Y | |
| Number of Exposures (single-epoch images) in cCoadd | 16,248 | 16,117 | 15,792 | 15,731 | 12,329 | ... |
| Single-epoch Median PSF FWHM (arcseconds) | 1.11 | 0.95 | 0.88 | 0.83 | 0.90 | ... |
| Single-epoch Median Sky Brightness (mag arcsec$^{-2}$) | 22.05 | 21.18 | 19.92 | 18.74 | 17.97 | ... |
| Single-epoch Median Effective Image Noise (mag arcsec$^{-2}$)[a] | 25.25 | 24.95 | 24.32 | 23.59 | 22.31 | ... |
| Single-epoch Median Astrometric Repeatability (angular distance, milliarcseconds) | 34 | 28 | 29 | 34 | 53 | Section 4.1 |
| Single-epoch Photometric Repeatability (mmag) | 2.4 | 2.1 | 2.2 | 1.8 | 2.7 | Section 4.2.1 |
| Single-epoch Magnitude Limit (PSF, S/N = 10)[b] | 23.57 | 23.34 | 22.78 | 22.10 | 20.69 | Morganson et al. 2018 |
| Single-epoch Median Saturation Limit (point sources, mag)[b] | 15.2 | 15.7 | 15.8 | 15.5 | 13.6 | Section 3 |
| Coadd Sky Coverage (individual bands, deg$^2$) | 5055 | 4988 | 5023 | 5031 | 5080 | Section 4.4 |
| Coadd Sky Coverage (grizY union, deg$^2$) | | | 4913 | | | Section 4.4 |
| Coadd Median Astrometric Relative Precision (angular distance, milliarcseconds) | | | 27 | | | Section 4.1 |
| Absolute Photometric Uncertainty (stat., mmag)[c] | 2.2 | 2.2 | 2.2 | 2.2 | 3.3 | Section 4.2.2 |
| Absolute Photometric Uncertainty (sys., mmag)[d] | 11 | 11 | 11 | 12 | 12 | Section 4.2.2 |
| Coadd Photometric Precision (mmag) | ≲2.4 | ≲2.1 | ≲2.2 | ≲1.8 | ≲2.7 | Section 4.2.1 |
| Coadd Photometric Uniformity versus Gaia (mmag) | | | 2.15 | | ... | Section 4.2.1 |
| Coadd Magnitude Limit (MAG_APER_4, 1″.95 diameter, S/N = 10) | 24.7 | 24.4 | 23.8 | 23.1 | 21.7 | Section 4.5 |
| Coadd 95% Completeness Limit (mag) | 24.6 | 24.3 | 24.0 | 23.7 | 23.4 | Section 4.5 |
| Coadd Spurious Object Rate | | | ≲1% | | | Section 4.6 |
| Coadd Galaxy Selection (EXTENDED_COADD ⩾ 2, 19.0 ⩽ mag_auto_i ⩽ 22.5) | Efficiency > 99%; Contamination > 2% | | | | | Section 4.7 |
| Coadd Stellar Selection (0 ⩽ EXTENDED_COADD ⩽ 1, 19.0 ⩽ mag_auto_i ⩽ 22.5) | Efficiency > 94%; Contamination < 3% | | | | | Section 4.7 |

**Notes.**
[a] Square root of the calibrated image variance, including read noise.
[b] Single-epoch magnitude and saturation limits in the Y band are provided for 45 s exposures.
[c] The given passband's MAGERR_AUTO added in quadrature with the estimated "Coadd Photometric Uniformity versus Gaia" (2.15 mmag).
[d] The absolute value of the difference between the MAG_AUTO AB offset in the given passband estimated from c26202_stisnic_007 and that estimated from the DA white dwarf sample, divided by 2. Because the DA white dwarf sample only provides estimates of the AB color offsets, an i-band offset estimated from the newer c26202_stiswfcnic_002 was assumed for the DA sample in order to estimate AB magnitude offsets for all passbands (see Table 2). We note that the absolute photometric systematic uncertainties quoted here simply reflect the difference in the C26202- and the DA-based methods of absolute color calibration and do not result from nonuniformities in the DES DR2 data. The nonuniformities are considered in the absolute photometric statistical errors.

CCD with a typical full-well depth starts to saturate for point sources observed at the median seeing and exposure time (90 s for griz and 45 s for Y) of the DES wide-field survey at about $g = 15.2$, $r = 15.7$, $i = 15.8$, $z = 15.5$, and $Y = 13.6$ mag. A summary of data quality metrics for the single-epoch data is reported in Table 1.

The multiepoch pipeline produces coadded images and catalogs of astronomical objects organized within a tiling scheme that subdivides the full celestial sphere into manageable-sized regions for parallel processing (Morganson et al. 2018). Each coadd "tile" is a square region of $0°.7306 \times 0°.7306$ with adjacent tiles overlapping by $\sim 30″$. Coadded images are created in each tile with dimensions of 10,000 pix × 10,000 pix and a pixel scale of 0″.263 (approximately the native pixel scale of DECam) using a sinc interpolation (LANCZOS3). Tile images are rescaled such that the magnitude ZP is fixed to 30 for all filters and the values of the pixels are given in units of picomaggies.[72] This makes the conversion between flux and magnitude the same for all bands.

The set of input single-epoch images are based on data quality assessments from the Final Cut pipeline and FGCM. In addition, single-epoch images with severe scattered light, ghosting, or bright transient defects (e.g., comets, meteors, airplanes, and earthquakes) are rejected. Input single-epoch images are astrometrically aligned to each other and to an external reference catalog from Gaia DR2 using SCAMP (Bertin 2006). After astrometric alignment, SWarp (Bertin et al. 2002; Bertin 2010) is used to form coadded images and a detection coadded image from the linear combination of the $r + i + z$ coadd images. PSFEx is then used to obtain a PSF model for each tile (Bertin 2011). Initial catalogs are constructed using SourceExtractor in double-image mode where the detection image is used to form the segmentation map of sources prior to extracting measurements from the tiles in each band.

We caution that the coadd PSF model is unable to fully account for discontinuities that occur at single-epoch image boundaries, thus limiting the precision of measurements of coadd PSF quantities to no better than a few percent (DES Collaboration 2018b). To overcome this limitation, the single-epoch catalogs are matched to the coadd detection catalog and weighted averages of the single-epoch MAG_PSF and SPREAD_MODEL measurements are made from all unflagged detections of the same object for each band. These weighted-average measurements are prefixed by WAVG_ and are included in the public data release products. More details are given in Appendix A.

---
[72] http://www.sdss3.org/dr8/algorithms/magnitudes.php

Multiepoch, multiband, and multiobject fitting that overcomes the PSF discontinuities has been applied for DES cosmology analyses and will be released in data products associated with those publications.

### 3.1. Changes in Single-epoch Processing

Below we highlight several changes in the DES DR2 Final Cut processing of individual DECam exposures relative to the processing described in Morganson et al. (2018).

1. New image calibration products (e.g., master biases, super flats, pupil corrections) were created for the additional epochs of DES data taking. The process for creating these products is described in Section 3 and Appendix A of Morganson et al. (2018).
2. Gaia DR2 (Gaia Collaboration 2016) was used for the external astrometric reference catalog. This yielded improved absolute calibration as documented in Section 4.1.
3. ZP for all DES single-epoch images were rederived using FGCM (Burke et al. 2018). Because the FGCM relies on overlapping exposures to derive relative ZP, DES DR2 improves the accuracy and uniformity of ZP derived for exposures included in DES DR1 (see Section 4.2).
4. Starting in DES Year 5, a "light bulb" defect (a pixel with large spurious charge production) manifested itself at $x, y = 795, 2620$ on CCD = 46 (detector N15). The amplitude of this feature was found to depend on the state of the camera and the sky level of the image. An algorithm was used to detect the extent of this feature and set BPM_BADPIX over the affected region (MSK = BADPIX_BPM = 1).
5. Increased charge transfer inefficiency was detected sporadically in Amp B of CCD 41 (detector N10) during data taking in DES Year 6. An algorithm was developed to identify and mask affected pixels on this amplifier. The BPM_BADAMP bit was set for these pixels in the mask plane (MSK = BADPIX_BADAMP = 8).
6. The algorithm for detecting and masking artificial satellites was adjusted to avoid masking large, nearby galaxies. This algorithm was also improved to interpolate and extrapolate satellite trails over the focal plane, leading to increased detection efficiency.
7. To improve PSF model fitting, the PSFEx configuration was modified to use subsampled PSF models (0.7 × the native pixel size) and roughly twice the angular extent (11.″2 compared to 6.″6 for DR1), using more stable centroids (WIN_IMAGE) and more conservative object rejection (removing objects with any masked/flagged pixels).
8. The SourceExtractor source detection threshold was lowered to detect fainter sources (the single-epoch source detection threshold now approaches $3\sigma$).

### 3.2. Changes in Multiepoch Processing

The coadded image production and processing pipelines for DES DR2 closely follow those described in Morganson et al. (2018). We describe minor changes to this pipeline below.

1. The criterion for constructing coadded images in a tile was altered to require that $>50\%$ of the tile area was covered in all bands ($g, r, i, z, Y$) with a depth of $\geqslant 3$ exposures per band. (In DR1 the criterion was that $>50\%$ of the tile was covered in the $r, i, z$ bands to a depth of $\geqslant 2$ exposures per band.) This reduces the number of tiles from 10,338 in DES DR1 to 10,169 in DES DR2. The stricter requirement increases survey uniformity, but results in a slight reduction in survey area around the border of the DES footprint (e.g., compare the coadd tiles shown in the S82 region of Figure 1 to Figure 2 of DES Collaboration 2018b).
2. The rejection of single-epoch images with bright transients (e.g., comets, meteors, and airplanes) was accomplished by a rapid, by-eye scan of focal plane postage stamps. The rejection of single-epoch images with scattered light and ghosting used a ray-tracing algorithm and the positions of bright stars to determine a subset of images that were examined more closely (again by eye). All images were re-examined in preparation for DES DR2, resulting in slight differences from DES DR1. With more observations in DES DR2, the tendency has been to be stricter with the data quality because individual single-epoch images are less critical to the coverage. When coupled with the new coadd construction criteria (see above) these changes have resulted in a few "holes" in the DES coverage, which arise due to the rejection of images with scattered light artifacts and higher backgrounds in the bluer bands around bright stars (e.g., $\alpha$ Pav, $\alpha$ Gru, $\beta$ Gru, $\alpha$ Eri, and $\alpha$ Col; Burke et al. 2018; Sevilla-Noarbe et al. 2021).
3. Gaia DR2 is used as the external astrometric reference catalog for SCAMP (Bertin 2006). This results in a significant improvement in the astrometric accuracy relative to DES DR1, which used 2MASS as the astrometric reference catalog (Skrutskie et al. 2006). See Section 4.1 for more details on the astrometric calibration accuracy.
4. The SWarp background subtraction module can create artifacts in the vicinity of bright, extended objects (e.g., large galaxies, clusters, and bright stars). In DR2, an additional set of experimental coadd images is constructed using SWarp without applying local background subtraction (SUBTRACT_BACK = N). These images rely solely on the single-epoch pipeline full-focal-plane background subtraction model to remove background emission and do provide cosmetically improved images in the vicinity of bright extended objects that may be better suited for measurements in such cases. No DR2 catalog measurements originate from these images.
5. The procedure for using SWarp to combine the $r + i + z$ coadded images to produce a detection image was changed from COMBINE_TYPE CHI − MEAN to COMBINE_TYPE AVERAGE, as it was found to produce a more robust detection of faint objects in the presence of diffuse emission than our previous configuration.
6. The PSFEx configuration was updated in a manner similar to the single-epoch configuration. While these changes improve PSF performance, it does not solve the discrete nature of PSF variation in the tiles.
7. The SourceExtractor configuration has been altered to lower the source detection threshold. The detection limit in the object catalog is now ∼$5\sigma$ (compared to a detection threshold of ∼$10\sigma$ in DES DR1).

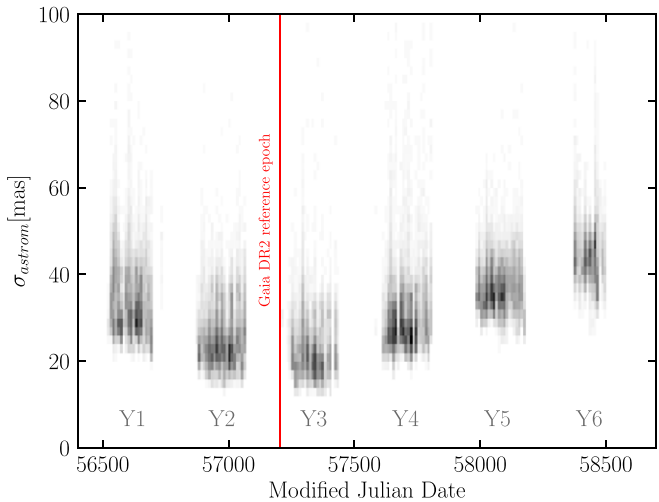

**Figure 3.** Astrometric precision per single-epoch image, calculated as the standard deviation of the 2D astrometric residuals with respect to the reference Gaia DR2 catalog. All bands are included.

## 4. Data Quality

The stability and sensitivity of DECam/Blanco combined with the DES observing strategy and processing pipeline provide an unparalleled deep, wide-area photometric data set. In this section, we provide a general assessment of the DES DR2 data quality including astrometric and photometric precision, survey footprint, imaging depth in terms of measured signal-to-noise and detection completeness, and morphological object classification accuracy. A summary of data quality metrics is reported in Table 1.

### 4.1. Astrometry

The DR2 astrometric solution is derived in two steps using SCAMP with Gaia DR2 as the reference catalog (see Morganson et al. 2018 for a description of the process). The internal astrometric uncertainties at the single-epoch level are $g = 34$, $r = 28$, $i = 29$, $z = 34$, and $Y = 53$ mas, as determined from the median of two-dimensional angular separations between repeated measurements of bright stars from individual exposures. In Figure 3, we show the standard deviation of the 2D astrometric residuals for each single-epoch exposure with respect to the Gaia DR2 reference catalog. The relative astrometry is most precise for exposures taken near the Gaia DR2 reference epoch. The trend across the six DES observing seasons is likely due to stellar proper motions, which are not accounted for in the DES DR2 astrometric solution. More precise relative astrometry can be achieved through additional modeling as described in Bernstein et al. (2017).

Prior to coaddition, an astrometric refinement step uses SCAMP to perform a simultaneous astrometric fit to all images contributing to a given coadd tile (all bands are considered simultaneously). The internal astrometric precision is estimated from the standard deviation of the residuals from these fits. The median of this value over all coadd tiles in DES DR2 is ∼27 mas.

### 4.2. Photometry

The DES DR2 photometry is validated in several ways including photometric uniformity, absolute calibration relative to reference standards, and the response to interstellar extinction. In the process of validating DES DR2, we have transformed several other surveys—SDSS (Albareti et al. 2017), PanSTARRS1 (Flewelling et al. 2020), HSC-SSP (Aihara et al. 2019), CFHTLenS (Erben et al. 2013), and a compilation of Johnson–Cousins photometry (Stetson 2009; Stetson et al. 2019).[73] These photometric transformation equations are provided in Appendix B.

#### 4.2.1. Relative Calibration

The internal, or relative, photometric solution for DES DR2 is derived with the FGCM approach (Burke et al. 2018). The FGCM fits model parameters using calibration stars observed in exposures taken during photometric conditions, as described in Appendix D in the reference above. The same network of stars is then used to refine the calibration of exposures taken in nonphotometric conditions. This approach was demonstrated to achieve $< 3$ mmag precision for point sources in the DES Y3 data (Sevilla-Noarbe et al. 2021).

The DES DR2 single-epoch photometric precision (associated with statistical and systematic uncertainty in the FGCM fit parameters) derived from repeated measurements of FGCM calibration stars is $g = 2.4$, $r = 2.1$, $i = 2.2$, $z = 1.8$, and $Y = 2.7$ mmag. These single-epoch photometric repeatability measurements include chromatic corrections and other local per-CCD gray corrections determined from the full FGCM model.

The internal photometric precision of the coadd is expected to be better than the single-epoch repeatability measurements because the statistical uncertainty can be reduced by combining fit results from multiple observations of the same fields. However, gains in the coadd photometric uniformity are only realized to the extent that these successive observations are uncorrelated and yield statistically independent model fit parameters. We anticipate that unknown systematic effects imprint correlations between model fits, though the degree of correlation is challenging to estimate from the DES data alone. Accordingly we quote the single-epoch repeatability as an upper bound on the uncertainty of the coadd relative photometry.

As a validation of the photometric uniformity, we compare the DES DR2 photometry to the photometry from the space-based Gaia DR2 $G$ band across the DES footprint (Figure 4). $G_{\text{pred}}$ is predicted from DES DR2 magnitudes by random forest regression trained on a sample of FGCM standard stars matched to Gaia DR2, using DES $r$, $g - r$, $r - i$, and $i - z$ as input parameters (Sevilla-Noarbe et al. 2021). Specifically, variations in relative uniformity are found to be $\sigma = 2.15$ mmag, as estimated from a Gaussian fit to the distribution of differences between $G_{\text{pred}}$ and $G_{\text{Gaia}}$. This uncertainty represents the difference between two completely independent sets of measurements, and therefore the internal dispersion of each survey must be smaller.

#### 4.2.2. Absolute Calibration

The DES DR2 absolute calibration is tied to the AB magnitude system (Oke & Gunn 1983; Fukugita et al. 1996) based on DES observations of the Hubble Space Telescope (HST) CALSPEC standard star C26202 (Bohlin et al. 2014)—in particular, to the spectrophotometry of the HST CALSPEC optical/NIR spectrum `c26202_stisnic_007`.[74] This follows the same procedure as was used in DES DR1 (DES Collaboration 2018b); however, several caveats should be noted.

---

[73] https://www.canfar.net/storage/list/STETSON/Standards

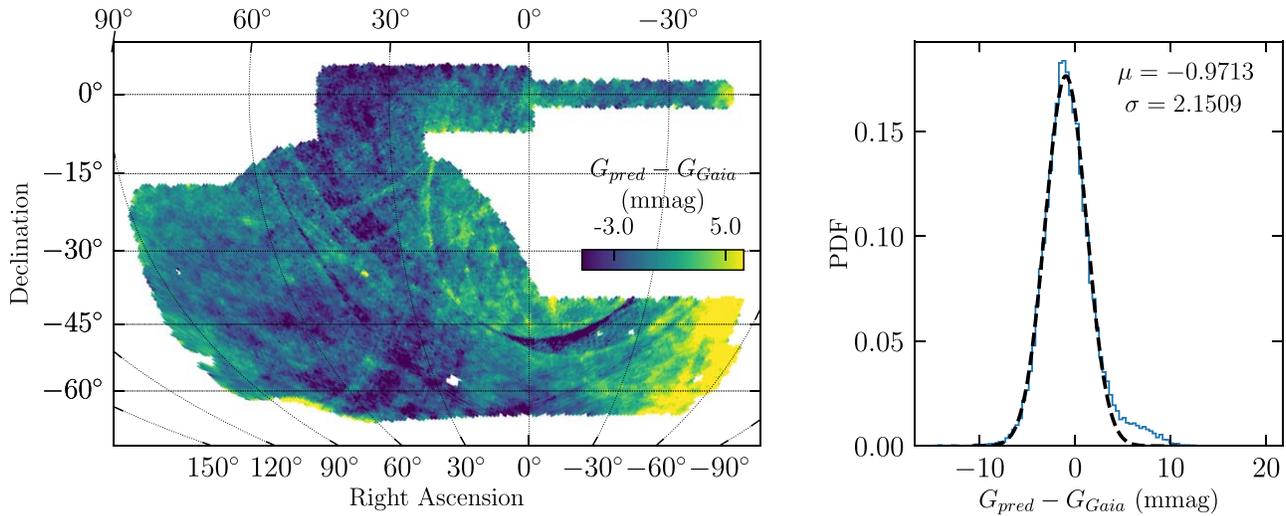

**Figure 4.** The photometric uniformity of the DES DR2 calibration can be estimated from the distribution of photometric residuals compared to the Gaia *G* band. Left: mean value of this residual vs. sky position in nside = 256 `HEALPix` pixels. Note that the most prominent features in this map are associated with the Gaia DR2 observing strategy. Right: normalized histogram of photometric residuals over the footprint fit by a Gaussian (dashed line) with mean and standard deviation quoted in the top-right corner.

First, the *AB* offsets were calculated based on a source catalog of individual measurements of C26202 calibrated by the FGCM. This calibration was performed before image coaddition for each of these two data releases, and subtle differences in the selection of "good" observations of C26202 (generally based on flux repeatability tests in the single-epoch data and/or on the choice of image processing flags to apply) and the flux measurement algorithm used can yield differences at the few millimagnitude level in the absolute calibration for different magnitude measurements, such as `WAVG_MAG_PSF` and `MAG_AUTO`. Thus, there can be small, but nonzero, differences (at the few millimagnitude level) between C26202's DES DR2 magnitudes and its FGCM *AB*-calibrated magnitudes (e.g., see the offsets for `c26202_stisnic_007` listed in Table 2).

Second, an updated and larger sample of *AB* reference stars is now available, including an updated version of the HST CALSPEC spectrum (`c26202_stiswfcnic_002`) as well as a "Golden Sample" of ∼150 relatively faint ($16.5 \lesssim r \lesssim 18$) pure-hydrogen-atmosphere ("DA") white dwarfs in the DES footprint (Figure 5).

In Table 2, we provide *AB* offsets that can be applied to the DES DR2 `WAVG_MAG_PSF` and `MAG_AUTO` magnitudes and colors to place them more closely onto the *AB* magnitude system (for details, see Appendix C). For other DES magnitude types, please see the online documentation. We also report two contributions to the uncertainty of the *AB* offsets: (1) the internal precision (statistical error) that indicates how well the DES photometry is tied to a given *AB* reference standard (`c26202_stisnic_007`, `c26202_stiswfcnic_002`, or the DA white dwarf sample), and (2) the external accuracy (systematic error) that indicates how well the particular reference standard is itself tied to the *AB* system.

The internal precision or statistical error is primarily driven by observational error. Thus, in the case of C26202, we calculate the statistical errors for each passband by adding in quadrature the Poisson-based magnitude errors from the DES DR2 catalog (e.g., `MAGERR_AUTO_I`) for C26202 with the estimated coadd uniformity error from the previous section (2.15 mmag). For example, for `MAG_AUTO` *i*-band, this would be

$$\sigma_{\text{stat}}(i) = \sqrt{(0.42 \text{ mmag})^2 + (2.15 \text{ mmag})^2} = 2.19 \text{ mmag}, \quad (2)$$

where `MAGERR_AUTO_I` = 0.42 mmag. This is how we calculated the statistical uncertainty in the absolute photometric calibration for *g*, *r*, *i*, *z*, *Y* reported in Table 1 [75] (and for the C26202 *i* band in Table 2).

For the statistical error in the colors of C26202, we assume that the nonuniformity in the relative calibrations is perfectly correlated between filter bands; so the 2.15 mmag uniformity uncertainty cancels out and we only calculate the statistical error based on the Poisson-based magnitude errors of the two filters in question. For example, for `MAG_AUTO` $i - z$, this would be

$$\sigma_{\text{stat}}(i-z) = \sqrt{(0.42 \text{ mmag})^2 + (0.58 \text{ mmag})^2} = 0.72 \text{ mmag}, \quad (3)$$

where `MAGERR_AUTO_I` = 0.42 mmag and `MAGERR_AUTO_Z` = 0.58 mmag. This is how we calculated the statistical errors for C26202 in $g - r$, $r - i$, $i - z$, and $z - Y$ in Table 2.

For the statistical errors for the DA white dwarf *AB* color offsets, because the reference stars sparsely sample the entire DES footprint, we use the scatter in the values from the ≈150 stars to estimate the precision, and these values are listed in Table 2.

---

[74] https://archive.stsci.edu/hlsps/reference-atlases/cdbs/calspec/

[75] In Table 1, we list the `c26202_stisnic_007` values for $\sigma_{\text{stat}}$, because those reflect the absolute photometric precision of the DES DR2 data as provided. Further, we list the `MAG_AUTO` values there; the `WAVG_MAG_PSF` values are similar.

**Table 2**
Photometric Offsets (in Magnitudes) from the AB Magnitude System for the Two Main Magnitude Definitions Used in DES DR2, MAG_AUTO and WAVG_MAG_PSF

| AB Reference Standard | Band/Color | | | | |
|---|---|---|---|---|---|
| | $i$ | $g-r$ | $r-i$ | $i-z$ | $z-Y$ |
| | MAG_AUTO | | | | |
| c26202_stisnic_007[a] | | | | | |
| AB offset | −0.0010 | +0.0025 | −0.0016 | −0.0054 | +0.0069 |
| $\sigma_{\rm stat}$ | ±0.0022 | ±0.0005 | ±0.0005 | ±0.0007 | ±0.0026 |
| $\sigma_{\rm sys}$ | ±0.011 | ±0.0066 | ±0.0054 | ±0.0076 | ±0.0130 |
| c26202_stiswfcnic_002[b] | | | | | |
| AB offset | −0.0066 | +0.0001 | −0.0046 | -0.0021 | +0.0160 |
| $\sigma_{\rm stat}$ | ±0.0022 | ±0.0005 | ±0.0005 | ±0.0007 | ±0.0026 |
| $\sigma_{\rm sys}$ | ±0.011 | ±0.0053 | ±0.0040 | ±0.0092 | ±0.0084 |
| DES DA White Dwarfs[c] | | | | | |
| AB offset | …[d] | −0.0106 | −0.0125 | −0.0205 | +0.0328 |
| $\sigma_{\rm stat}$ | … | ±0.0013 | ±0.0012 | ±0.0012 | ±0.0011 |
| $\sigma_{\rm sys}$ | … | ±0.0053 | ±0.0040 | ±0.0092 | ±0.0084 |
| | WAVG_MAG_PSF | | | | |
| c26202_stisnic_007[a] | | | | | |
| AB offset | −0.0010 | +0.0014 | −0.0012 | −0.0058 | +0.0009 |
| $\sigma_{\rm stat}$ | ±0.0022 | ±0.0005 | ±0.0005 | ±0.0005 | ±0.0014 |
| $\sigma_{\rm sys}$ | ±0.011 | ±0.0062 | ±0.0040 | ±0.0072 | ±0.0126 |
| c26202_stiswfcnic_002[b] | | | | | |
| AB offset | −0.0066 | −0.0010 | −0.0042 | −0.0025 | +0.0100 |
| $\sigma_{\rm stat}$ | ±0.0022 | ±0.0005 | ±0.0005 | ±0.0005 | ±0.0014 |
| $\sigma_{\rm sys}$ | ±0.011 | ±0.0049 | ±0.0025 | ±0.0088 | ±0.0081 |
| DES DA White Dwarfs[c] | | | | | |
| AB offset | …[d] | −0.0109 | −0.0091 | −0.0201 | +0.0261 |
| $\sigma_{\rm stat}$ | … | ±0.0013 | ±0.0007 | ±0.0006 | ±0.0006 |
| $\sigma_{\rm sys}$ | … | ±0.0049 | ±0.0025 | ±0.0088 | ±0.0081 |

**Notes.** For other magnitude definitions like MAG_APER and MAG_PETRO, please visit the online documentation at https://des.ncsa.illinois.edu/releases/dr2. Offsets for the DES DR2 MAG_AUTO and WAVG_MAG_PSF magnitudes. These are not applied by default in the DES DR2 photometry. They should be applied as mag$_{\rm AB}$ = mag$_{\rm DR2}$ + offset (or, likewise, color$_{\rm AB}$ = color$_{\rm DR2}$ + offset).
[a] The Hubble CALSPEC standard spectrum used for the nominal absolute calibration of DES DR1 and DES DR2.
[b] The most recent version of the Hubble CALSPEC standard spectrum for C26202.
[c] These offsets were determined based on a "Golden Sample" of ~150 DA white dwarfs that were spectroscopically observed and modeled explicitly for purpose of the absolute color calibration of the DES. (Modeling was performed using the DA white dwarf models described in Tremblay et al. 2011, 2013.)
[d] For those who may wish to make use of the AB offsets derived from the DA white dwarfs, an AB offset needs to be adopted for the $i$ band. For this case, adopting the $i$-band offset from c26202_stiswfcnic_002 is currently recommended.

For rough estimates of the external accuracy, or systematic uncertainty, we compare the C26202-based results against our DA-based results. Due to the interplay between how DA models help define the CALSPEC system and how the CALSPEC system helps define the absolute calibration of ground-based spectra (like those of the DES DR2 DA white dwarfs), these two samples are unlikely to be completely independent. We report our best estimates of the systematic uncertainty based on the information at hand, but we note that the true systematic uncertainty is likely slightly larger.

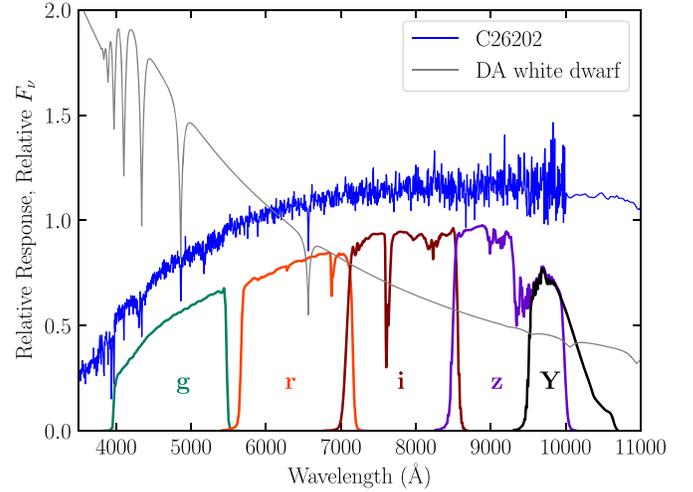

**Figure 5.** A plot of the DES Standard Bandpass curves ($grizY$; see also Figure 1 of DES Collaboration 2018b) superposed by the spectrum for the HST CALSPEC spectrophotometric standard C26202 (c26202_stiswfcnic_002) and the model spectrum of one of the stars in the DES DA white dwarfs "Golden" sample. (This particular DES DA white dwarf has $T_{\rm eff}$ = 29, 087 ± 357 K, $\log g$ = 8.169 ± 0.067). The "break" in the c26202_stiswfcnic_002 spectrum at $\lambda \approx 10000$ Å is due to differing characteristics of the HST STIS spectrograph used for the spectrum at $\lambda \lesssim 10000$Å and those of the HST NICMOS2 grisms used for the spectrum at $\lambda \gtrsim 10000$Å.

In Table 2, for the $g-r$, $r-i$, $i-z$, and $z-Y$ color offsets, we estimate the systematic uncertainties by taking the absolute value of the difference between the C26202 result and the DA white dwarf result and then dividing by 2 (we do this separately for the c26202_stisnic_007-based results and for c26202_stiswfcnic_002-based results.) For the $i$-band magnitude, we have no independent measures of the offsets from our DA sample. Therefore, we estimate the systematic uncertainties in the $i$-band AB offset by adding in quadrature the estimated uncertainty in tying C26202 to the CALSPEC system as a whole (≈1% [≈10 mmag] in the optical; R. C. Bohlin 2021, private communication; Bohlin et al. 2019) with the estimated uncertainty in tying the CALSPEC system itself to the AB system (≈0.5% [≈5 mmag] in the optical; Bohlin 2014; Bohlin et al. 2020), i.e.,

$$\sigma_{\rm sys}(i) = \sqrt{(10\ {\rm mmag})^2 + (5\ {\rm mmag})^2} \approx 11\ {\rm mmag}. \quad (4)$$

We likewise calculate the systematic uncertainties in the AB offsets for the other bands ($g$, $r$, $z$, $Y$) tabulated in Table 1, but increasing the uncertainty in how well the CALSPEC is tied to the AB system from 5 to 7 mmag for the $z$ and $Y$ bands (based in part on Section 2.3 and Figure 4 of Bohlin et al. 2020).

Additional details of the absolute calibration of the DES DR2 can be found in Appendix C and in a forthcoming paper.

*4.2.3. Interstellar Reddening*

The DR2_MAIN table includes a column for $E(B-V)$ values from the reddening map of Schlegel et al. (1998) (SFD98) at the location of each catalog object. Following DES DR1, the $E(B-V)$ values were obtained using a linear interpolation of the Zenithal Equal Area projected map distributed by SFD98. The FGCM magnitudes can be corrected by an amount

Table 3
Depth Estimates for DES DR2

| Method | Band | | | | |
| --- | --- | --- | --- | --- | --- |
| | g | r | i | z | Y |
| Maximum in number counts (MAG_AUTO) | 25.0 | 24.7 | 24.3 | 24.0 | 23.2 |
| Measured with S/N = 10 (MAG_AUTO) | 24.0 | 23.8 | 23.1 | 22.3 | 20.7 |
| Measured with S/N = 10 (MAG_APER_4) | 24.7 | 24.4 | 23.8 | 23.1 | 21.7 |
| Imaging depth from mangle (MAG_APER_4) | 24.7 | 24.4 | 23.8 | 23.1 | 21.7 |
| Detection completeness of 95% (MAG_AUTO) | 24.6 | 24.3 | 24.0 | 23.7 | 23.4 |

**Note.** Depth estimates for DES DR2 after selecting objects with FLAGS_[GRIZY] < 4 and IMAFLAGS_ISO_[GRIZY] = 0. The imaging depth from mangle has been obtained directly from the images based on the coaddition of single-epoch images. The detection completeness is estimated with respect to HSC-PDR2 deep fields.

$A_b = E(B-V)R_b$, where $R_b$ is computed per band as described in DES Collaboration (2018b) using the DES DR1 Standard Bandpasses, which remain valid. Following Section 4.2 of DES Collaboration (2018b), we apply the calibration adjustment of Schlafly & Finkbeiner (2011) to our fiducial reddening coefficients so that these coefficients can be used directly with $E(B-V)$ values from the SFD98 reddening map. The values for $R_b$ are thus the same as for DES DR1 and are replicated here for completeness: $R_g = 3.186$, $R_r = 2.140$, $R_i = 1.569$, $R_z = 1.196$, and $R_Y = 1.048$. The photometric measurements included in the DES DR2 database tables are not dereddened by default. However, the DR2_MAIN table includes additional columns for the dereddened versions of MAG_AUTO and WAVG_MAG_PSF indicated by a _DERED suffix (Appendix E).

### 4.3. Flagged Objects

The identification and flagging of catalog objects for DES DR2 are the same as in DES DR1. Two different flags are included for each object in the catalog in each band g, r, i, z, Y separately:

1. IMAFLAGS_ISO: indicates if any pixel within the isophotal area of the object was flagged in all the individual images that comprise the coadded image of the object (Morganson et al. 2018). This occurs most frequently near bright stars, bleed trails, and gaps in coverage near the edge of the footprint. We recommend the use IMAFLAGS_ISO = 0 in all bands to select a clean sample of objects. For a complete description of the flag bits, see the online documentation.
2. FLAGS: includes the standard SourceExtractor flags, which are the same as DES DR1 (see Table D1). For more details, see the online SourceExtractor documentation.[76] In this case, we generally recommend FLAGS < 4, which does not reject blended objects or objects close to bright neighbors.

### 4.4. Footprint

As part of DES DR2, we distribute HEALPix (Górski et al. 2005) maps of the individual g, r, i, z, and Y image coverage, with spatial resolution comparable to the size of gaps between individual CCDs (nside = 4096, ∼0″.86; Drlica-Wagner et al. 2018). In addition, we provide HEALPix indices for all objects at different resolutions in the DR2_MAIN table.

The creation of these maps follows the method described in Section 7.1 of Drlica-Wagner et al. (2018). The DESDM processing creates maps of the survey coverage at the coadded image level including masked regions around bright stars, bleed trails, and artificial satellite streaks using mangle (Hamilton & Tegmark 2004; Swanson et al. 2008). These maps are then translated into high-resolution HEALPix maps, where the value of each pixel represents the fraction of the pixel area that is covered by DES DR2. The coverage fraction (FRACDET) ranges from 0 to 1, where a value of 1 indicates that the pixel has complete, unmasked coverage. File format and access details are given in Section 5. An example of a g-band coadded image on the border of the footprint and the associated coverage fraction map is shown in Figure 6. The map captures the lack of DES observations, as well as areas that are masked due to bright stars and other imaging artifacts.

The area of the DES DR2 footprint can be calculated by summing the coverage fraction maps in each band individually. This yields an estimated footprint area in g, r, i, z, Y = 5055, 4988, 5023, 5031, and 5080 deg² (see Table 1). We estimate the area that is covered in all five bands by summing the minimum coverage fraction in each pixel with at least one exposure in $grizY$.[77] This yields an estimated coverage of 4913 deg². The area of DES DR2 is approximately 200 deg² smaller than that of DES DR1 due to stricter image rejection criteria and a more restrictive requirement on the definition of the coverage while coadding (Section 3). In Figure 7, we show the cumulative distribution of the footprint area as a function of the fractional pixel area. Approximately 97% of the DES DR2 area has a coverage fraction of >80%.

### 4.5. Depth

We derive several conventional depth estimates based on the DES DR2 images and catalogs. These estimates are provided as a standard reference that can be compared to other surveys and previous DES data releases. In general, the depth will depend on the object sample, and we recommend that users re-evaluate the depth for their specific object sample. The DES DR2 catalogs extend deeper than DES DR1 due to the larger number of images going into the coadds and the lower source detection threshold.

We perform five different estimates of the depth. For the catalog-based estimates in this section, we use all catalog objects without performing any star/galaxy selection. First, we determine the mode of magnitude distributions of cataloged objects using MAG_AUTO; second and third, from the magnitude corresponding to a fixed S/N (= 10), both for MAG_AUTO and MAG_APER_4 (equivalent to a 1.95″ aperture); fourth, we use the mangle information to estimate the magnitude at which S/N = 10 within the same aperture; and fifth, we estimate the object detection completeness with respect to the HSC-SSP PDR2 deep fields. The only pre-selection we apply for these tests are the recommended quality flags cuts: FLAGS_[GRIZY] < 4 and IMAFLAGS_ISO_[GRIZY] = 0. Results from each estimate are provided in Table 3.

---

[76] https://sextractor.readthedocs.io/en/latest/Flagging.html

[77] A more rigorous estimate of the multiband coverage area will be provided with future DES cosmology analyses.

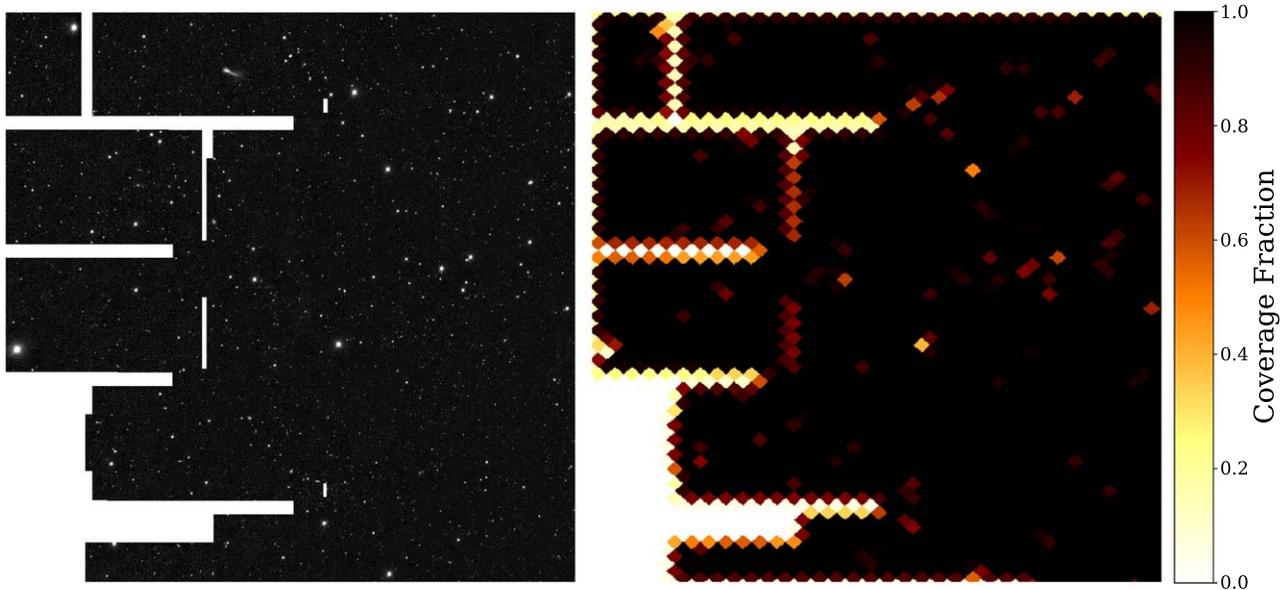

**Figure 6.** DES *g*-band coadded image (left) and coverage fraction map (right) for tile DES 0301−0458 located at the border of the DES footprint.

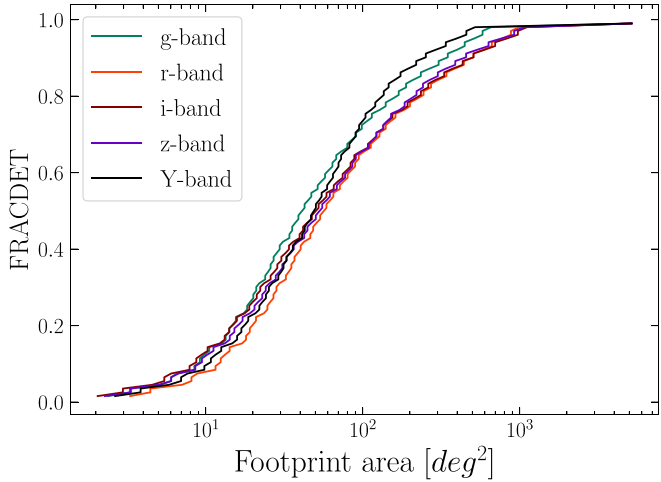

**Figure 7.** Cumulative footprint area as a function of the fractional pixel coverage (FRACDET) on a logarithmic scale. In each band separately, if we select pixels with FRACDET > 0.8 we eliminate ∼130 deg$^2$ of the survey area.

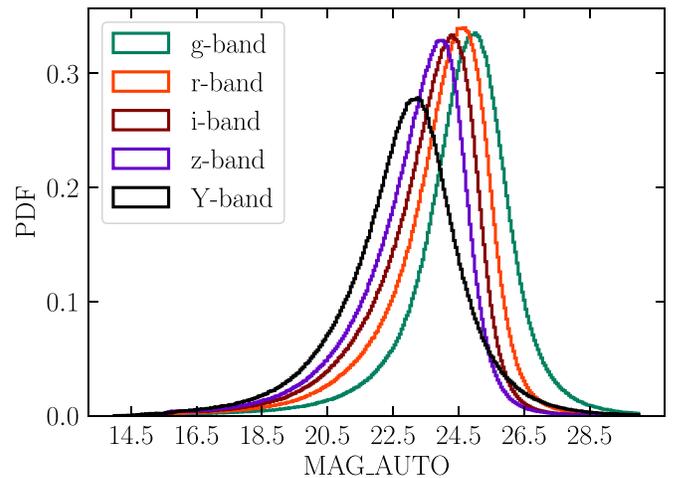

**Figure 8.** Normalized histograms, *n*(mag), of sources as a function of SourceExtractor's MAG_AUTO quantity. No signal-to-noise selection has been applied. A coarse estimate of depth can be derived from the mode of these distributions. See the first row in Table 3 for values.

DESDM performs object detection on an $r + i + z$ detection coadd and measures object properties in each band using SourceExtractor in "double-image mode" (Section 3). This method increases the detection efficiency in each band and results in catalogs that contain many faint objects measured with low S/N in the shallower bands (i.e., *z* and *Y*). This causes appreciable differences between depth estimates based solely on object detection and those that impose a fixed S/N cut. As expected, the largest differences between these estimates can be seen in the *z* and *Y* bands (Table 3).

*4.5.1. Flux Distribution*

We can infer a crude estimate of the detection limit assuming that the intrinsic number of astronomical objects will increase at fainter magnitudes. Therefore, the completeness threshold can be estimated as the magnitude where the number of objects stops increasing. We calculated the number of catalog objects as a function of magnitude, *n*(mag), and estimate the mode of the distribution to estimate the survey depth. Using MAG_AUTO, the mode of the distribution is $g = 25.0$, $r = 24.7$, $i = 24.3$, $z = 24.0$, and $Y = 23.2$. This represents an increase in magnitude limit of $\Delta(g, r, i, z, Y) \sim 0.7, 0.8, 0.9, 1.9, 1.8$ mag with respect to DES DR1. The *n*(mag) distribution can be found in Figure 8. Note that no signal-to-noise cut has been applied, therefore these values should be used with caution.

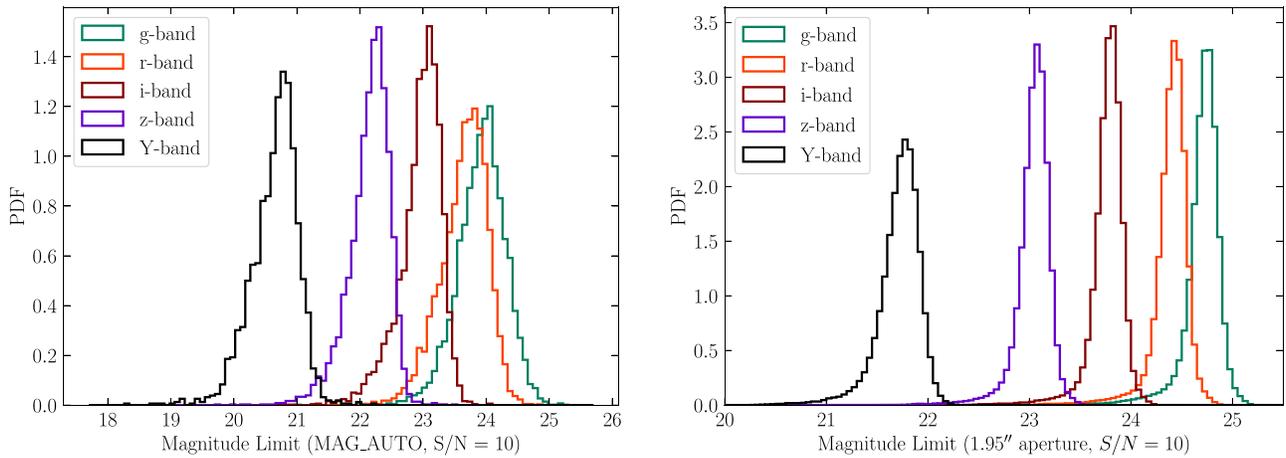

**Figure 9.** Two estimates of the DES DR2 coadd catalog depths. Left: magnitude distribution for sources with S/N = 10 (MAGERR_AUTO = 0.10857). The median value of each distribution indicates the depth. Right: catalog depth for a 1″.95 aperture estimated from image properties using mangle. Values are given in Table 3.

### 4.5.2. Magnitude Limit at Fixed Signal-to-noise Ratio

The depth for a given S/N-limited sample can be determined from the distribution of magnitude uncertainties as a function of magnitude (e.g., Rykoff et al. 2015). The statistical magnitude uncertainty, $\delta m$, is related to the S/N, $F/\delta F$, via Pogson's law (Pogson 1856) and propagation of uncertainties,

$$\delta m = \frac{2.5}{\ln 10} \frac{\delta F}{F}. \quad (5)$$

We use this formula to estimate the depth of DES DR2 for sources with S/N = 10 ($\delta m \approx 0.10857$). We estimate depths both for MAG_AUTO and MAG_APER_4 by selecting sources with 0.10837 < MAGERR < 0.10877 and measuring the median magnitude value. We show histograms of the MAG_AUTO magnitudes of objects satisfying this criterion in the left panel of Figure 9. For MAG_AUTO we obtain an S/N = 10 magnitude limit of $g = 24.0$, $r = 23.8$, $i = 23.1$, $z = 22.3$, and $Y = 20.7$. This corresponds to an increase in magnitude limit of $\Delta(g, r, i, z, Y) \sim 0.6, 0.7, 0.6, 0.5$, and 0.2 mag with respect to DES DR1. For MAG_APER_4 we obtain an S/N = 10 magnitude limit of $g = 24.7$, $r = 24.4$, $i = 23.8$, $z = 23.1$, and $Y = 21.7$. This corresponds to an increase in magnitude limit of $\Delta(g, r, i, z, Y) \sim 0.4, 0.3, 0.4, 0.4$, and 0.3 mag with respect to DES DR1. These magnitude limits are given in the second and third rows of Table 3.

### 4.5.3. Depth from Image Properties

During image reduction, we compute coadd weight maps from the weighted sum of the single-epoch input images, using mangle (Hamilton & Tegmark 2004; Swanson et al. 2008). This weight is then converted to an S/N = 10 limiting magnitude for a 2″ diameter aperture, corresponding approximately to the MAG_APER_4 quantity measured by SourceExtractor (for details, see Drlica-Wagner et al. 2018).

From these maps, we can estimate the median limiting magnitude across the DES footprint. For DES DR2 these are $g = 24.7$, $r = 24.4$, $i = 23.8$, $z = 23.1$, and $Y = 21.7$ (right panel of Figure 9). This corresponds to a magnitude limit increase of $\Delta(g, r, i, z, Y) \sim 0.4, 0.5, 0.5, 0.4$, and 0.4 mag with respect to DES DR1.

### 4.5.4. Object Detection Completeness

We can also assess the effective depth by comparing to deeper imaging data. In this case, we evaluate the depth of DES DR2 data with respect to the Hyper-Suprime Camera PDR2 deep/udeep fields (HSC-PDR2; Aihara et al. 2019), using the overlap in the DEEP2-F3 field centered on (R.A., decl.) = (352°.5, 0°.4) and in the SXDS field, centered on (R.A., decl.) = (35°.5, −5°.4). The 5σ depth for point sources in the HSC-PDR2 deep fields is $g$, $r$, $i$, $z$, and $y \sim 27.3, 26.9, 26.7, 26.3$, and 25.3 mag, as measured in the HSC system (Aihara et al. 2019). The overlap area is ∼18 deg².

Before matching the HSC and DES catalogs, we apply the DES DR2 footprint masks to both surveys, selecting pixels with full coverage in all DES bands and then we apply the HSC bright star mask. Next, we clean the DES DR2 sample with the recommended quality flags cuts.

The HSC-PDR2 sample is significantly deeper than DES DR2 and thus suffers more severely from the blending of objects (Bosch et al. 2018; Melchior et al. 2018). In our analysis, we identified a large number of unmatched sources coming from blending. To avoid this issue, we only select isolated sources in HSC by removing all sources that are within 5″ of each other. Finally, we apply a 10σ cut in the HSC $g$, $r$, $i$, $z$, and $y$ magnitudes.

We match the HSC-PDR2 catalog to DES DR2 with a matching radius of 1″. The DES DR2 detection efficiency is defined as the fraction of HSC-PDR2 objects in a given magnitude interval (expressed in the DES photometric system after converting magnitude systems; Appendix B) that have a matched object in DES DR2.

The detection efficiency curves are plotted in Figure 10. The 95% completeness magnitude limit is $g = 24.6$, $r = 24.3$, $i = 24.0$, $z = 23.7$. and $Y = 23.4$ (fifth row in Table 3). This corresponds to a detection efficiency increase of $\Delta(g, r, i, z) \sim 0.9, 1.0, 1.1$, and 1.4 mag with respect to DES DR1.

### 4.6. Spurious Object Rate

The increased depth and decreased detection threshold of DES DR2 have several implications for source detection and background modeling. While the dominant impact is to improve the catalog completeness for faint objects, there may be differences in measurement and deblending that do not exactly match those for DES DR1. Given the increased number of observations, $N$, there is a linear increase in the overall number of artifacts, though the effective brightness of most artifacts in the coadded images is reduced by roughly $\sqrt{N}$. Artifacts include unmasked cosmic rays, unmasked artificial satellite and space debris trails, unmasked bleed trails and edge-bleed artifacts, interloping solar system objects, and scattered light artifacts.

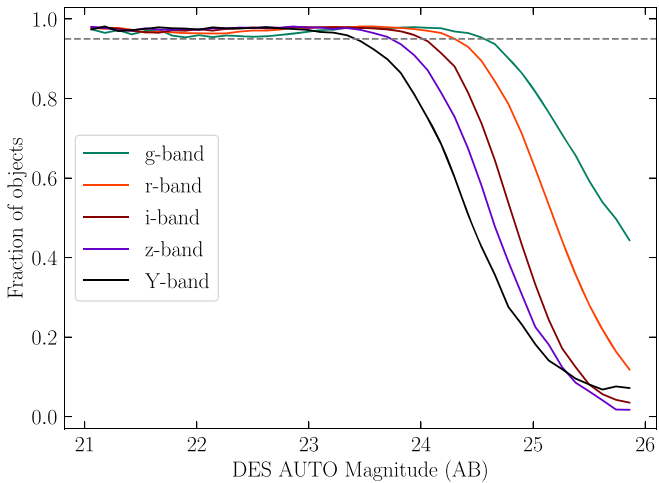

**Figure 10.** DES DR2 detection efficiency with respect to the Hyper-Suprime Camera PDR2 deep fields. No signal-to-noise selection has been applied, and measurements with negative flux in DES DR2 data have been removed. For visual reference, the gray dashed line indicates 95% of objects.

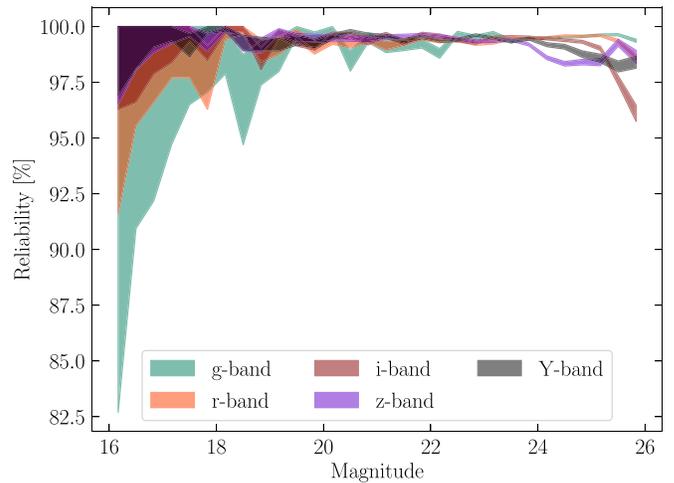

**Figure 11.** Reliability (complement of the spurious detection rate after visual inspection) of the DES DR2 catalog as a function of magnitude. The reliability is calculated through comparison with the DES Y3 Deep Fields (Hartley et al. 2020), and errors in the reliability are calculated following the Bayesian technique described in Paterno (2004). The total fraction of spurious sources is $\lesssim 1\%$.

We estimate the "reliability" (complement of the spurious detection rate) of the DES DR2 object catalog by comparing it to the DES Y3 Deep Fields catalog (Hartley et al. 2020), which is ~1.3 mag deeper. We select sources in the DES DR2 sample that are not present in the Deep fields, after matching both catalogs using a 1″ radius. We find that ~1.3% of sources are unmatched. However, most of these sources are real physical objects with large apparent movements (e.g., solar system objects) or short-duration transient sources. To estimate the fraction of truly spurious sources, we visually inspect a subsample of the unmatched objects (~2300) and find that ~34% are actually spurious. Our investigation confirms that the spurious object contamination level is $\lesssim 1\%$, similar to DES DR1. In Figure 11, we plot the reliability (defined as the complement of the spurious detection rate after visual inspection) as a function of magnitude.

### 4.7. Morphological Object Classification

DES DR2 includes a morphological object classification variable, EXTENDED_COADD, to delineate samples of spatially extended galaxies and samples of point-like stars and quasars (e.g., Sevilla-Noarbe et al. 2021). EXTENDED_COADD is based on the SourceExtractor SPREAD_MODEL quantity that compares the relative fit quality between the local PSF model and a slightly extended circular exponential disk model convolved with the PSF (Desai et al. 2012). As shown in the top panel of Figure 12, stars are located in a tight locus with SPREAD_MODEL near zero, whereas galaxies have positive SPREAD_MODEL values.

To support science cases with different completeness and purity requirements, we combine a set of three Boolean conditions to define samples of high-confidence stars (EXTENDED_COADD =0), likely stars (EXTENDED_COADD =1), likely galaxies (EXTENDED_COADD =2), and high-confidence galaxies (EXTENDED_COADD =3). The value of EXTENDED_COADD is defined as

```
EXTENDED_COADD =
 ((SPREAD_MODEL_I + 3 SPREADERR_MODEL_I) > 0.005)
+((SPREAD_MODEL_I + SPREADERR_MODEL_I) > 0.003)
+((SPREAD_MODEL_I − SPREADERR_MODEL_I) > 0.002),
```
(6)

where each conditional statement that is true increases the value of EXTENDED_COADD by one. For example, users can select benchmark galaxy (EXTENDED_COADD $\geqslant$ 2) and stellar samples (0 $\leqslant$ EXTENDED_COADD $\leqslant$ 1) having 543 million and 145 million objects, respectively, after applying a basic object quality selection (IMAFLAGS_ISO=0 and FLAGS < 4 in the $i$ band). The baseline classifier implementation above uses coadd measurements in the $i$ band. Similar approaches could use SPREAD_MODEL in other bands (e.g., Palmese et al. 2016), or the weighted average of SPREAD_MODEL measurements from individual exposures (e.g., Drlica-Wagner et al. 2015). The classification performance of WAVG_SPREAD_MODEL and SPREAD_MODEL is similar for bright objects (DES Collaboration 2018b).

Approximately 0.03% of all objects have invalid SPREAD_MODEL_I values. These objects are assigned sentinel values SPREAD_MODEL_I = 0 and SPREADERR_MODEL_I = 0 (objects with *imaflags_iso*[$i$] = 1; see Section 4.3) or SPREAD_MODEL_I = 1 and SPREADERR_MODEL_I = 1 (failed fit). We flag these objects in the benchmark morphological classifier as EXTENDED_COADD = −9.

The bottom two panels of Figure 12 show the classification accuracy of EXTENDED_COADD evaluated relative to HSC-SSP DR1 (Aihara et al. 2018). We consider three overlap regions spanning the HSC SXDS (Ultra Deep layer), DEEP2_3 (Deep layer), and VVDS (Wide layer) fields covering ~18 deg$^2$. The depth and excellent image quality of the HSC-SSP ($i$-band seeing FWHM $\lesssim 0\farcs7$) enable robust tests of the DES DR2 morphological classifier to $i \sim 24$, reaching the

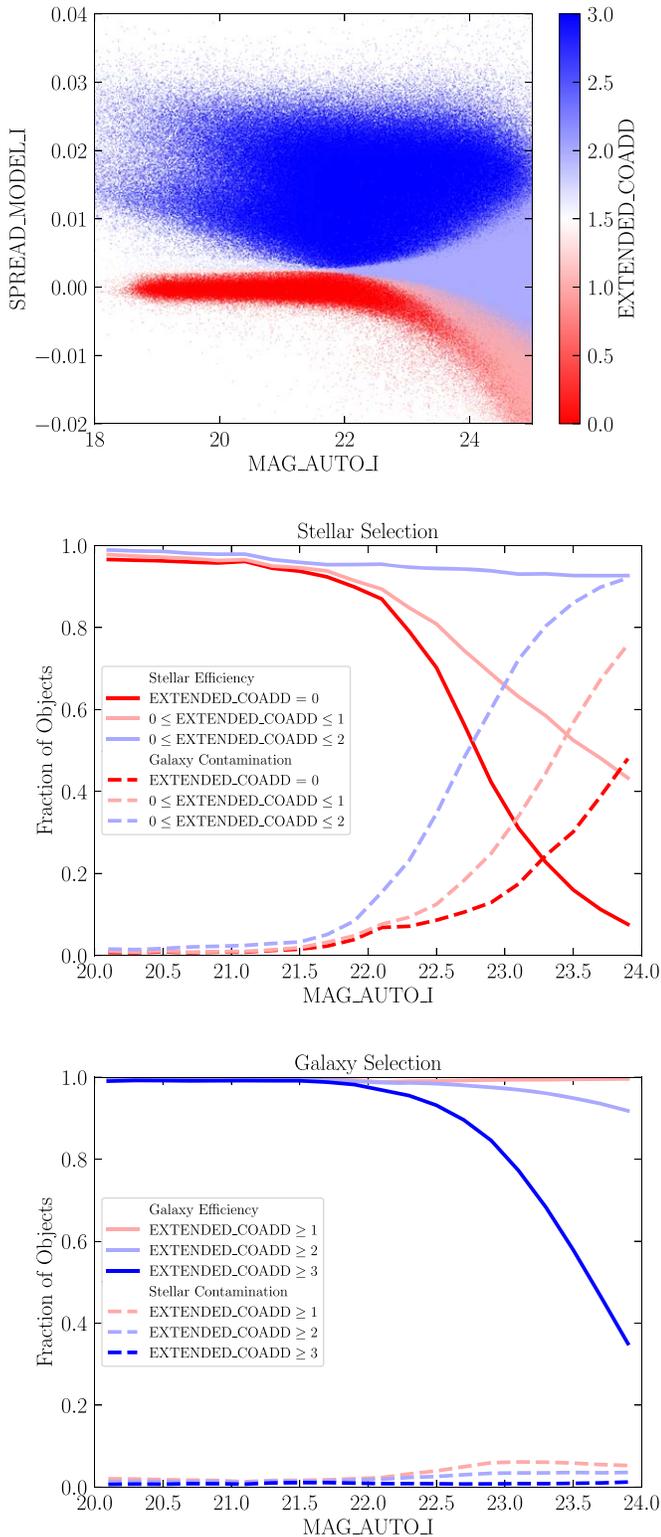

**Figure 12.** The EXTENDED_COADD morphological classifier defines samples of point-like and spatially extended objects based on their coadd SPREAD_MODEL and SPREADERR_MODEL values (top). The loci of stars and galaxies are well separated at bright magnitudes and partially overlap at faint magnitudes. The two lower panels show the classification accuracy evaluated with respect to HSC-SSP for several intervals of EXTENDED_COADD values for stellar (middle) and galaxy (bottom) samples.

95% detection completeness threshold of DES DR2. Table 4 summarizes the classification efficiency and purity for star and galaxy samples measured with respect to HSC-SSP.

**Table 4**
Summary of Morphological Classification Efficiency and Purity for the Baseline DES DR2 Classifier EXTENDED_COADD for Star and Galaxy Samples in Two Magnitude Ranges

| Quantity | Efficiency | Purity |
|---|---|---|
| Stellar Selection: 19. <MAG_AUTO_I <22.5 | | |
| EXTENDED_COADD $= 0$ | 0.9239 | 0.9770 |
| $0 \leqslant$ EXTENDED_COADD $\leqslant 1$ | 0.9417 | 0.9703 |
| $0 \leqslant$ EXTENDED_COADD $\leqslant 2$ | 0.9711 | 0.9286 |
| Galaxy selection: 19. <MAG_AUTO_I <22.5 | | |
| EXTENDED_COADD $\geqslant 1$ | 0.9934 | 0.9774 |
| EXTENDED_COADD $\geqslant 2$ | 0.9913 | 0.9825 |
| EXTENDED_COADD $\geqslant 3$ | 0.9775 | 0.9912 |
| Stellar selection: 19. <MAG_AUTO_I <23.5 | | |
| EXTENDED_COADD $= 0$ | 0.7700 | 0.9571 |
| $0 \leqslant$ EXTENDED_COADD $\leqslant 1$ | 0.8615 | 0.8864 |
| $0 \leqslant$ EXTENDED_COADD $\leqslant 2$ | 0.9609 | 0.5745 |
| Galaxy selection: 19. <MAG_AUTO_I <23.5 | | |
| EXTENDED_COADD $\geqslant 1$ | 0.9934 | 0.9579 |
| EXTENDED_COADD $\geqslant 2$ | 0.9790 | 0.9738 |
| EXTENDED_COADD $\geqslant 3$ | 0.8648 | 0.9915 |

Spatial maps of the number counts of high-confidence stars and galaxies are shown in Figure 13. The Fornax and Sculptor dwarf spheroidal galaxies, the periphery of the LMC, the Sagittarius stellar stream, and increasing stellar density toward the Galactic plane are visible in the stellar map, whereas the galaxy map shows the cosmic web of large-scale structure.

### 4.8. Known Issues

We discuss known features of DES DR2, starting with those that are shared with DES DR1. As in DES DR1, the effective PSF for coadded images exhibits sharp discontinuities at the edges of overlapping single-epoch images that are not fully captured by a smoothly varying coadd PSF model. More accurate coadded photometry is achieved by using PSF models from individual images to either compute a weighted average of single-epoch measurements or to perform a joint multiepoch fit (e.g., Drlica-Wagner et al. 2018; Sevilla-Noarbe et al. 2021). Likewise, we observe correlated spatial variations in the difference between the coadded PSF and weighted-average PSF photometry (MAG_PSF − WAVG_MAG_PSF) at the level of ∼0.1 mag attributed to coadded PSF modeling. Accordingly, for point-source photometry, we recommend the use of WAVG_MAG_PSF for bright sources and MAG_AUTO for studies including faint sources.

The SPREAD_MODEL quantity used for morphological object classification also depends on the coadd PSF model. For bright stars, residuals between WAVG_SPREAD_MODEL and WAVG_SPREAD_MODEL are tightly correlated with residuals between the coadd PSF and weighted-average PSF photometry, and a significant part of the width of the coadd SPREAD_MODEL distribution for high-S/N stars is likely due to imperfections in coadd PSF modeling. More substantial coadd PSF fit failures occur in $\lesssim 0.1\%$ of the footprint. Multiepoch photometry for faint objects fitted with the PSF model of individual exposures is provided by ngmix in value-added DES data releases (e.g., Sevilla-Noarbe et al. 2021). Photometric residuals between PSF and aperture magnitudes at the few millimagnitude level are observed for FGCM photometric calibration stars that appear to be correlated with the local object density (Eckert et al. 2020).

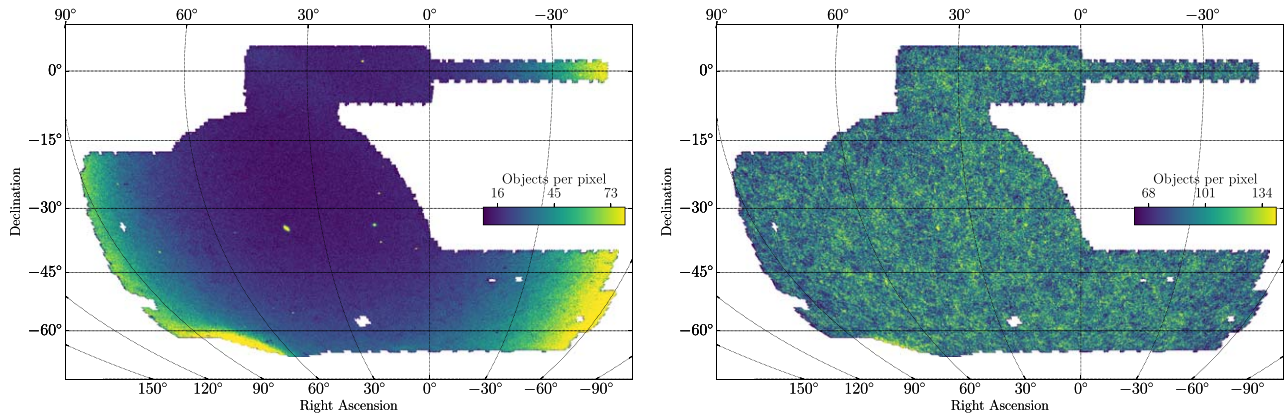

**Figure 13.** Left: map of stellar number counts created with the EXTENDED_COADD = 0 selection described in Section 4.7. Discrete peaks in the stellar number counts correspond to globular clusters and dwarf galaxies in the Milky Way halo. Right: analogous galaxy number counts map created with the EXTENDED_COADD = 3 selection. Color range units are the number of objects per HEALPix nside = 1024 pixel without correcting for the coverage fraction of each pixel. Both maps use a magnitude threshold of MAG_AUTO_I < 23.

A total of ∼0.5% of the sources have been flagged by IMAFLAGS_ISO in at least one of the *grizY* bands, and < 0.1% have artifacts in all five bands. As described in Section 4.3 it is recommended to use *imaflags_iso* = 0 as the first filter of clean objects.

Reflected and scattered light from very bright stars impacts the photometry of nearby objects. Exclusion regions around extremely bright stars have been enlarged compared to DES DR1 through more stringent image rejection criteria and more restrictive requirements on the number of images for coadd creation (Section 3.2). However, we expect spurious objects associated with bright stars to pass the source-extraction algorithm. We recommend the identification and removal of these sources at the catalog level through a set of quality cuts (e.g., Drlica-Wagner et al. 2018; Sevilla-Noarbe et al. 2021).

The DESDM processing pipeline is designed for extragalactic science. As such, it is not optimized to detect or measure sources in extremely crowded regions. Blending can be severe near the cores of Milky Way globular clusters, the centers of bright Milky Way satellite galaxies, the periphery of the Large Magellanic Cloud, and semiresolved nearby galaxies. Furthermore, the survey depth is now sufficient that emission from Galactic cirrus is visible across much of the footprint.

Associating DES DR2 objects with a given temporal epoch is not straightforward. The coadded images are assembled from single-epoch images collected over six years of observations. The MJD − OBS keyword in the coadded image headers represents the MJD of the first single-epoch image that went into that coadded image and is not representative of the epoch of source measurements. The astrometric calibration of DES DR2 is performed against Gaia DR2 sources reported at the J2015.5 epoch in the Gaia-CRF2 frame, which is well matched to the ICRF3 prototype to within 20 to 30 μas (Gaia Collaboration 2018). However, proper motions were not included in the SCAMP astrometric calibration, resulting in larger astrometric uncertainty for exposures taken in the years before and after the Gaia DR2 reference epoch (see Figure 3).

Finally, compared to DES DR1, the data comprising DES DR2 have received less internal use by the DES Collaboration at the time of this release. We expect that the online documentation will be periodically updated with any new guidance and recommendations for users.[78]

## 5. Release Products and Data Access

This section briefly summarizes the data products and data access tools for DES DR2.

### 5.1. Images

DES DR2 images can be grouped into two categories:

1. Calibrated Single-epoch Images: DES DR2 includes a total of 96,263 exposures with photometric calibration, corresponding to 5,796,641 individual calibrated single-epoch CCD images. The number of exposures per band is $N_g = 20{,}043$, $N_r = 20{,}106$, $N_i = 19{,}545$, $N_z = 22{,}713$, and $N_Y = 13{,}856$. There are 83,706 exposures coming from the DES wide-area survey, 76,217 of which passed quality cuts and were effectively used for building the coadded images. An additional 12,557 exposures released in DES DR2 come from other DES operations programs. Each exposure is ∼0.5 Gbyte in size (compressed). The processed CCD images for these exposures can be accessed through Astro Data Lab.

2. Coadded Images: This collection consists of 111,859 coadded images associated with 10,169 tiles of 0.534 deg² in size. For each tile, DESDM creates five background-subtracted images in *grizY*, five background-included images in *grizY*, and a single detection image coming from the combination of the $(r + i + z)$ background-subtracted coadded images. The coadded source catalogs are based on the background-subtracted images. Each individual image is ∼0.2 Gbyte in size (compressed) and are accessed through the NCSA DESaccess interface.[79]

Images are compressed using the fpack utility (Pence et al. 2010).

---
[78] https://des.ncsa.illinois.edu/releases/dr2
[79] https://des.ncsa.illinois.edu/desaccess/

### 5.2. Catalogs and Database Tables

The coadd source detection process cataloged 691,483,608 distinct objects. Object information includes object centroids, shape parameters, `HEALPix` indices, and processing flags. Several different photometric measurements and associated uncertainties are provided, including `AUTO`, `PETRO`, `WAVG_PSF`, and assorted aperture magnitudes (Table D2). These measurements are distributed in three database tables: `DR2_MAIN`, `DR2_MAGNITUDE`, and `DR2_FLUX` served from an Oracle database at NCSA. The `DR2_MAIN` table contains all object information that is not a photometric measurement or uncertainty, augmented by `MAG_AUTO` and `WAVG_MAG_PSF` (with associated uncertainties) and interstellar-extinction-corrected versions of these magnitudes. The other two tables contain additional magnitude and flux measurements (with associated uncertainties). Coordinates, flags, and `HEALPix` indices are present in all three tables. We note that they contain two sets of coordinates for the objects, (R.A., decl.) and (`ALPHAWIN_J2000`, `DELTAWIN_J2000`)[80] that are computed in the same manner; the difference between them is that `RA` and `DEC` are truncated to six decimals to facilitate indexing and table partitioning on these columns, while `ALPHAWIN_J2000` and `DELTAWIN_J2000` are double-precision quantities to be used when precise measurements are needed. All spatial-based queries should use `RA` and `DEC` in their condition statements.

The table `DR2_TILE_INFO` contains information about the processed tiles, such as sky location, image geometry, number of objects, and file paths to the associated images and object catalogs. Finally, `DR2_COVERAGE` contains the footprint `HEALPix` indices with their fractional coverage in each band. For a complete description of these tables, we refer the reader to Appendix E.

### 5.3. Files

We have created FITS file versions of the `DR2_MAIN`, `DR2_MAGNITUDE`, and `DR2_FLUX` tables grouped by coadd tile. This amounts to 30,507 total files with almost 3 Tbyte of catalog data. Both the catalog and image file paths can be obtained from the `DR2_TILE_INFO` table (see online documentation for example queries) and can be accessed through the interfaces described in Section 5.7.

### 5.4. Coverage Maps

DR2 includes high-resolution `HEALPix` maps of the survey footprint in units of partial coverage (see Section 4.4). The fractional coverage within each spatial pixel is calculated separately for each of the $grizY$ bands. The maps are distributed at resolution nside = 4096, corresponding to a pixel side length of ∼0.′86. The coverage maps distributed in FITS file format contain two columns, `PIXEL` and `FRACDET`, and are saved as partial `HEALPix` maps for storage efficiency. They can be found on the DES DR2 release page, as well. Coverage maps can also be found in the database table `DR2_COVERAGE`.

### 5.5. DES Standard Bandpasses

DES DR2 uses the same standard bandpass as DES DR1. The Blanco/DECam total system response, including instrument and atmosphere, in the $grizY$ bands is available in digital form with the DECam online documentation.[81]

### 5.6. Software

All software used in the DESDM pipelines can be accessed from the release page itself[82] or from the DES GitHub Organization.[83] The configuration described in Morganson et al. (2018), together with access to the software used to generate the data products, are the main ingredients needed to reprocess the data in a manner similar to that done by DESDM. Note that in addition to the raw and processed single-epoch images, NOIRLab will provide long-term preservation, curation, and public access to DES DR1 and DES DR2.

### 5.7. Data Access

This section describes the tools and interfaces that provide access to DES DR2. Data access is supported by NCSA[84], LIneA[85], and Astro Data Lab.[86]

#### 5.7.1. NCSA DESaccess

NCSA continues to provide access to DES DR2 (and previous releases) with `DESaccess` (https://des.ncsa.illinois.edu/desaccess/), a collection of web applications connected to the Oracle database hosting the DES DR2 catalogs. The `DESaccess` API server and job management system were recently revamped to enhance the service and ensure its long-term sustainability. The user-facing web interface (UI) was also updated for compatibility with the new API and to add new features, including integrated online documentation of the available tools and a detailed specification of the `DESaccess` API to support programmatic use of the service. Here we present a brief description of the main services provided by `DESaccess`:

1. The DES Tables app provides a list of all tables the user can access. The user can also view descriptions of columns in the chosen schema.
2. The DB Access app provides an SQL web client that allows the user to send queries directly to the Oracle 12 database hosting the DES DR2 tables. Queries with a small result set (currently limited to 1000 lines) may be sent synchronously as a "quick query," where the results are displayed on the web page within a few seconds. Typically queries are submitted for asynchronous execution, with an option to notify the user by email when the results are available for download. Results can be retrieved in the following formats: `csv`, `FITS` (Wells et al. 1981), or `HDF5`.[87] Compression is optional for the `csv` and `HDF5` formats. The query interface is powered using `easyaccess` (Carrasco Kind et al. 2019),[88] an enhanced SQL command-line interpreter developed for astronomical surveys such as DES.

---

[80] We note that the astrometry of DES DR2 is actually tied to Gaia-CRF2 (Section 4.8).

[81] http://www.ctio.noao.edu/noao/content/DECam-filter-information
[82] https://des.ncsa.illinois.edu/releases/dr1/dr1-docs/processing
[83] https://github.com/DarkEnergySurvey
[84] http://www.ncsa.illinois.edu/
[85] http://www.linea.gov.br/
[86] https://datalab.noirlab.edu
[87] https://www.hdfgroup.org/solutions/hdf5/
[88] https://github.com/mgckind/easyaccess

3. The Cutout Service allows the user to request cutouts up to 12′ on a side centered on a given position, specified either by RA/DEC coordinates or coadd object ID. Cutout requests are highly configurable; output files may include image data from chosen bands in `FITS` format as well as detection images in PNG format generated using the `STIFF` (Bertin 2012) or the method of Lupton et al. (2004) from sets of color band triplets designating the red/green/blue color mapping. Similar to query jobs, cutout requests are processed asynchronously. The header data of the cutout files is a copy of the original header from the images with additional keys indicating the center of the cutout `RA_CUTOUT` and `DEC_CUTOUT`. No stitching is performed for objects near the edge of the tile.

4. The Tile Finder is a service for finding information on tiles, either by specifying the position on the sky the tile covers or by providing the tile name. Displayed information includes the tile name, center position, corner coordinates, and the number of objects. A full list of download links for catalogs and images associated with the tile is provided for each available data release.

User-generated job output files are stored for 10 days, with the option to extend storage up to 30 days. Query and cutout job information and results can be viewed on the "Job Status" page. Both the code powering `DESaccess`[89] and the DES release site are open sourced. The `DESaccess` website is built with HTML, JS, and Polymer[90] that provides a reusable web component framework. The `DESaccess` API server and job management system are written in Python and use packages including the Tornado web framework[91] and the Kubernetes Python API.[92] All of these applications are containerized using Docker[93] and run within a Kubernetes[94] cluster, providing a robust system enabling high-availability services with scalable deployments and efficient resource usage.

### 5.7.2. LIneA Science Server

The images and catalogs generated by DESDM for DES DR2 can also be accessed through a web interface developed by the Laboratorio Interinstitucional de e-Astronomia (LIneA) for DES DR1 and available at NCSA since early 2018.[95] A new landing page integrates the new APIs implemented at NCSA, including a new cutout service. The platform has been designed to offer different ways to visualize images, catalogs (from DES and other surveys), and target lists, which can be either user uploaded or created from queries to the database carried out using the platform's user query functionality. The platform includes the following services:

1. The Sky/Image Viewer combination integrates third-party tools to allow the user to visualize the entire sky map produced by DES (Aladin Lite[96] developed by CDS[97]) in the form of a HiPS (Hierarchical Progressive Surveys) image, as well as each individual coadd tile using `VisiOmatic` (Bertin et al. 2015). We have added the following functionalities to the native functions of Aladin: (1) a search for a position specified by the user; (2) the possibility of sharing a specific visualization (position and zoom) with other users; (3) a map viewer that allows the display of observational maps; (4) different grids and polylines including the original DES tiles and footprint border; and (5) access to the image of a specific DES tile, which can be examined in a separate screen using `VisiOmatic` and its native tools. The user can visualize a whole tile or inspect a specific position on the sky in more detail using the Image Viewer. The native functions of the tool include snapshots, profile overlays, contrast settings, color mix, zoom, full-screen mode, and catalog overlay. Other useful functions are switching on/off markers, recenter the display to visualize the entire tile, crop, and save part of the image. We have added several functionalities, which include: (1) the ability to download the `FITS` image and catalog associated with the tile being displayed; (2) the ability to perform a side-by-side comparison of the same region of the sky using different display settings or data releases; (3) the ability to share a position/region with another user as mentioned above; and (4) the ability to select the contrast from a set of predefined values. Finally, one can also overlay objects from public catalogs, which are classified into three categories: (1) targets—list of positions/objects either uploaded or created using the Target Viewer and User Query services described below; (2) object catalog—the DES DR1/DR2 object catalog produced by DESDM; and (3) external catalogs—a sample of catalogs available in Vizier.[98] The user can change the symbol, color, and size associated with each selected catalog to facilitate the comparison between catalogs. From the image viewer, it is possible to display catalog information for specific objects by clicking on a link that appears in a tooltip. This will redirect the user to a page containing all the available catalog information, a zoomed image of the object, the object's location in the DES footprint, and a low-resolution spectrum based on the flux in *grizY*. From this page, one can also access both the SIMBAD[99] and NED[100] services.

2. The Target Viewer enables the user to upload a list of objects or select objects using the User Query service described below. The first page provides a summary of the available target lists, the ability to upload, delete, and mark a list(s) as a favorite. The upload can currently be done manually by pasting lines of (R.A., decl.) coordinates to the interface or selecting a table in the user's database. In fact, the user can paste a list of (R.A., decl.) positions or a copy of the contents of a CSV file with a header (the first line) containing multiple columns, but with columns labeled R.A. and decl. (case insensitive). Future versions will allow files to be uploaded in `CSV` and `VOTable` format. Once a list is pasted or uploaded into the tool, the image surrounding an object/position can be visualized by selecting an entry in the list and the corresponding position. The user can then select the columns to be shown, sort according to a given attribute, and comment, rank, and reject an entry. One can also apply a filter, and the filtered list can be downloaded (as csv or FITS file) or saved for future use. This tool is useful, for instance, to analyze a list of candidates with specific properties (e.g., galaxies of a given morphological type, stars of a given spectral type, etc.). It is also possible to explore a given object's properties from this tool, as described above.

---
[89] https://github.com/des-labs
[90] https://www.polymer-project.org/
[91] https://www.tornadoweb.org/en/stable/
[92] https://github.com/kubernetes-client/python/
[93] https://www.docker.com/
[94] https://kubernetes.io/
[95] https://desportal2.cosmology.illinois.edu/
[96] http://aladin.u-strasbg.fr/AladinLite/
[97] http://cds.u-strasbg.fr/

[98] http://vizier.u-strasbg.fr/
[99] http://simbad.u-strasbg.fr/simbad/
[100] https://ned.ipac.caltech.edu/

3. The Tile Viewer is a simplified way to locate and visualize a specific DES tile or sky location.
4. The User Query service provides access to the database table available in the DES DR1/DR2 data from which SQL queries can be written, validated, and executed. The resulting table is displayed under "My Tables" where it can be renamed, a few lines of its content listed, and deleted. Objects selected can be immediately viewed in the Target Viewer after the resulting table columns are properly associated with those recognized by the tool.

Given the suite of functionalities available in each tool, video tutorials are available to introduce first-time users to the services.[101] To offer an independent DES DR2 data access center, LIneA is also preparing a mirror site with services completely hosted in Brazil, as we intend to provide long-term curation of the final DES data set.

### 5.7.3. Astro Data Lab

The Astro Data Lab (Fitzpatrick et al. 2016; Nikutta et al. 2020) science platform (https://datalab.noirlab.edu) is one of the access portals for DES DR2. It is being developed at the Community Science and Data Center (CSDC), which is hosted by NOIRLab. Astro Data Lab enables efficient exploration and analysis of large data sets including surveys that use DECam and the Mosaic cameras (Dey et al. 2016); high-value external photometric surveys; and, since recently, massively multiplexed spectroscopic surveys.

Astro Data Lab hosts the DES DR2 catalog tables using a PostgreSQL v13 database backend. The data are identical to those hosted by the DES Collaboration through the NCSA portal but with a few additions. First, the Astro Data Lab database contains tables with crossmatches to other large catalogs such as Gaia and AllWISE. Second, the main table contains several extra columns, such as ecliptic and Galactic coordinates, a Hierarchical Triangular Mesh (HTM)[102] index, supplementary HEALPix indices in RING (nside =256) and NESTED (nside =4096) schemes, S/Ns, and precomputed colors. Finally, the data are clustered and indexed using the Q3C scheme (Koposov & Bartunov 2006), which enables fast spatial queries. Astro Data Lab also serves the coadded images and single-epoch tiles, including a Simple Image Access (SIA) metadata database, and an image cutout service.

Access to DES DR2 data, to colocated auxiliary high-value data sets, and to various data services is possible in a number of ways:

1. Database access to catalogs. Facilitated through a Table Access Protocol (TAP)[103] service and via direct database queries. The query service recognizes both PostgreSQL and ADQL query languages. Both synchronous and asynchronous queries are supported, as well as anonymous and authenticated access. Queries can be launched either through a web-based interface or through TAP-aware clients such as TOPCAT[104] (Taylor 2005), or through the multipurpose datalab command-line utility,[105] or within Jupyter notebooks (both local and remote).
2. Simple Image Access and image cutout service. The Astro Data Lab Simple Image Access (SIA) service provides access to cutouts of the DES DR2 images, both coadds and single-epoch tiles. For a given position on the sky, an SIA service call returns a table of metadata of all compatible images. The metadata include select header information for each image and can be further subselected. A URL to retrieve a cutout of a specified size is also included.
3. Crossmatching service. Astro Data Lab provides a performant crossmatching service, both web-based and through Python APIs. Users can crossmatch any of their tables with any of the Data Lab holdings. Data Lab also serves precomputed crossmatch tables against standard astrometric, photometric, and spectroscopic surveys (for DES DR2: Gaia DR2 and EDR3, AllWISE, unWISE DR1, NOIRLab Source Catalog DR2, and SDSS DR16). These tables allow much faster spatial searches for counterparts (within the default 1″.5 radius).
4. File service. Access to all survey images (coadds and single-epoch files) and auxiliary DES survey files, as a file collection.
5. Remote compute/notebook server. Colocated with all data holdings, remote compute capabilities are furnished through a JupyterLab notebook analysis environment.[106] It allows the easiest mode of access to all data holdings and analysis tools, without the need to install any software at all or to download any data locally; only a web browser and an Astro Data Lab account are needed. All common astronomy and data science Python libraries are pre-installed, including the client interfaces specific to Data Lab (for queries, storage, access to a spectral service, and authentication).
6. Remote storage. Astro Data Lab provides users with space for files and database tables, enabling persistent notebooks and user data. We reserve 1 TB per user for file storage (VOSpace) and 250 Gbyte for personal DB tables (MyDB).
7. All-sky viewer based in Aladin Lite. The viewer supports multiple-layer transparency and hosts various wide-area surveys.
8. Curated set of example notebooks. The team at Astro Data Lab develops and curates a large collection of relevant example Jupyter notebooks for the users. The set features introductory notebooks, technical how-tos, complete science case examples, and education/public outreach and user-contributed notebooks. The notebooks are developed on Github[107] and we welcome pull requests.

---

[101] https://desportal2.cosmology.illinois.edu/tutorials
[102] http://www.skyserver.org/htm
[103] https://ivoa.net/documents/TAP
[104] The Astro Data Lab TAP URL is http://datalab.noirlab.edu/tap.
[105] https://github.com/noaodatalab/datalab
[106] Astro Data Lab Jupyter notebook server: https://datalab.noirlab.edu/devbooks.
[107] Astro Data Lab example notebooks: https://github.com/noaodatalab/notebooks-latest/.

More information on using the data services and our data holdings is available on the Astro Data Lab web page: https://datalab.noirlab.edu.

## 6. Summary

DES has completed a six-year imaging campaign covering ∼5000 deg$^2$ of the south Galactic cap with precise $grizY$ photometry to a depth of ∼ 24 mag at S/N = 10. The first major public release of DES data (DES DR1) was based on the first three seasons of DES observations (DES Collaboration 2018b). This second major public release (DES DR2) has approximately doubled the integrated exposure time in most areas of the footprint, increasing the photometric depth by 0.45–0.7 mag and raising the total number of cataloged objects from ∼400M to ∼700 M. Improvements have also been made in photometric calibration uniformity (< 3 mmag), internal astrometric precision (∼27 mas), and star–galaxy separation accuracy, as summarized in Table 1. The processed single-epoch images, coadded images, and coadded object catalogs that comprise DES DR2 are accessible via three complementary online platforms hosted at NCSA, LIneA, and NOIRLab.

While DES observations have concluded, the DES Collaboration continues to refine image reduction and object photometry algorithms and to produce new collections of value-added data products that will be released together with the legacy cosmological results. We anticipate that the wide-area catalog of DES DR2 will be complemented by a set of Deep Field data products assembled from the SN survey observations combined with community observations of selected fields (e.g., Hartley et al. 2020). The DES DR2 object catalog will be enhanced to produce the DES Y6 Gold catalog following a procedure similar to that used to produce DES Y3 Gold (Sevilla-Noarbe et al. 2021) from DES DR1. These and other DES data products and services used for a variety of science applications will be made publicly available at https://des.ncsa.illinois.edu/releases.

Funding for the DES Projects has been provided by the US Department of Energy, the US National Science Foundation, the Ministry of Science and Education of Spain, the Science and Technology Facilities Council of the United Kingdom, the Higher Education Funding Council for England, the National Center for Supercomputing Applications at the University of Illinois at Urbana-Champaign, the Kavli Institute of Cosmological Physics at the University of Chicago, the Center for Cosmology and Astro-Particle Physics at the Ohio State University, the Mitchell Institute for Fundamental Physics and Astronomy at Texas A&M University, Financiadora de Estudos e Projetos, Fundação Carlos Chagas Filho de Amparo à Pesquisa do Estado do Rio de Janeiro, Conselho Nacional de Desenvolvimento Científico e Tecnológico and the Ministério da Ciência, Tecnologia e Inovação, the Deutsche Forschungsgemeinschaft, and the Collaborating Institutions in the Dark Energy Survey.

The Collaborating Institutions are Argonne National Laboratory, the University of California at Santa Cruz, the University of Cambridge, Centro de Investigaciones Energéticas, Medioambientales y Tecnológicas-Madrid, the University of Chicago, University College London, the DES-Brazil Consortium, the University of Edinburgh, the Eidgenössische Technische Hochschule (ETH) Zürich, Fermi National Accelerator Laboratory, the University of Illinois at Urbana-Champaign, the Institut de Ciències de l'Espai (IEEC/CSIC), the Institut de Física d'Altes Energies, Lawrence Berkeley National Laboratory, the Ludwig-Maximilians Universität München and the associated Excellence Cluster Universe, the University of Michigan, NFS's NOIRLab, the University of Nottingham, The Ohio State University, the University of Pennsylvania, the University of Portsmouth, SLAC National Accelerator Laboratory, Stanford University, the University of Sussex, Texas A&M University, and the OzDES Membership Consortium.

Based in part on observations at Cerro Tololo Inter-American Observatory at NSF's NOIRLab (NOIRLab Prop. ID 2012B-0001; PI: J. Frieman), which is managed by the Association of Universities for Research in Astronomy (AURA) under a cooperative agreement with the National Science Foundation.

The DES data management system is supported by the National Science Foundation under Grant Numbers AST-1138766 and AST-1536171. The DES participants from Spanish institutions are partially supported by MICINN under grants ESP2017-89838, PGC2018-094773, PGC2018-102021, SEV-2016-0588, SEV-2016-0597, and MDM-2015-0509, some of which include ERDF funds from the European Union. IFAE is partially funded by the CERCA program of the Generalitat de Catalunya. Research leading to these results has received funding from the European Research Council under the European Union's Seventh Framework Program (FP7/2007-2013) including ERC grant agreements 240672, 291329, and 306478. We acknowledge support from the Brazilian Instituto Nacional de Ciência e Tecnologia (INCT) do e-Universo (CNPq grant 465376/2014-2).

This manuscript has been authored by Fermi Research Alliance, LLC under contract No. DE-AC02-07CH11359 with the US Department of Energy, Office of Science, Office of High Energy Physics.

This work made use of the Illinois Campus Cluster, a computing resource that is operated by the Illinois Campus Cluster Program (ICCP) in conjunction with the National Center for Supercomputing Applications (NCSA) and which is supported by funds from the University of Illinois at Urbana-Champaign.

This research is part of the Blue Waters sustained-petascale computing project, which is supported by the National Science Foundation (awards OCI-0725070 and ACI-1238993) and the state of Illinois. Blue Waters is a joint effort of the University of Illinois at Urbana-Champaign and its National Center for Supercomputing Applications.

The Science Server described here was developed and is operated by LIneA (Laboratório Interinstitucional de e-Astronomia).

We acknowledge support from the Australian Research Council through project numbers CE110001020, DP160100930, and FL180100168, and the Brazilian Instituto Nacional de Ciência e Tecnologia (INCT) e-Universe (CNPq grant 465376/2014-2).

This research uses services or data provided by the Astro Data Lab at NSF's National Optical-Infrared Astronomy Research Laboratory. NOIRLab is operated by the Association of Universities for Research in Astronomy (AURA), Inc. under a cooperative agreement with the National Science Foundation.

The Hyper Suprime-Cam (HSC) collaboration includes the astronomical communities of Japan and Taiwan, and Princeton

University. The HSC instrumentation and software were developed by the National Astronomical Observatory of Japan (NAOJ), the Kavli Institute for the Physics and Mathematics of the Universe (Kavli IPMU), the University of Tokyo, the High Energy Accelerator Research Organization (KEK), the Academia Sinica Institute for Astronomy and Astrophysics in Taiwan (ASIAA), and Princeton University. Funding was contributed by the FIRST program from the Japanese Cabinet Office, the Ministry of Education, Culture, Sports, Science and Technology (MEXT), the Japan Society for the Promotion of Science (JSPS), Japan Science and Technology Agency (JST), the Toray Science Foundation, NAOJ, Kavli IPMU, KEK, ASIAA, and Princeton University.

This paper makes use of software developed for the Large Synoptic Survey Telescope. We thank the LSST Project for making their code available as free software at http://dm.lsst.org.

The PanSTARRS1 Surveys (PS1) have been made possible through contributions of the Institute for Astronomy, the University of Hawaii, the Pan-STARRS Project Office, the Max-Planck Society and its participating institutes, the Max Planck Institute for Astronomy, Heidelberg, and the Max Planck Institute for Extraterrestrial Physics, Garching, The Johns Hopkins University, Durham University, the University of Edinburgh, Queen's University Belfast, the Harvard-Smithsonian Center for Astrophysics, the Las Cumbres Observatory Global Telescope Network Incorporated, the National Central University of Taiwan, the Space Telescope Science Institute, the National Aeronautics and Space Administration under grant No. NNX08AR22G issued through the Planetary Science Division of the NASA Science Mission Directorate, the National Science Foundation under Grant No. AST-1238877, the University of Maryland, and Eötvös Loránd University (ELTE), and the Los Alamos National Laboratory.

This work has made use of data from the European Space Agency (ESA) mission Gaia (https://www.cosmos.esa.int/gaia), processed by the Gaia Data Processing and Analysis Consortium (DPAC, https://www.cosmos.esa.int/web/gaia/dpac/consortium). Funding for the DPAC has been provided by national institutions, in particular the institutions participating in the Gaia Multilateral Agreement.

The authors would also like to thank R. C. Bohlin and S. E. Deustua for useful comments on the absolute calibrations sections.

*Facilities:* Blanco (DECam), Astro Data Lab.

*Software:* SourceExtractor (Bertin & Arnouts 1996), PSFEx (Bertin 2011), SCAMP (Bertin 2006), SWarp (Bertin et al. 2002; Bertin 2010), mangle (Hamilton & Tegmark 2004; Swanson et al. 2008), HEALPix (Górski et al. 2005),[108] healpy (Zonca et al. 2019),[109] healsparse,[110] matplotlib (Hunter 2007), numpy (van der Walt et al. 2011), scipy (Virtanen et al. 2020), astropy (Astropy Collaboration 2013), fitsio,[111] easyaccess (Carrasco Kind et al. 2019), skymap,[112] TOPCAT (Taylor 2005), fpack (Pence et al. 2010).

---

[108] http://healpix.sourceforge.net
[109] https://github.com/healpy/healpy
[110] https://healsparse.readthedocs.io/en/latest/
[111] https://github.com/esheldon/fitsio
[112] https://github.com/kadrlica/skymap

## Appendix A
## Weight-average Quantities

The weighted-average (WAVG) quantities are derived from the individual single-epoch SourceExtractor measurements for all detections of the same object in a given band that pass the following selection criteria: (1) FLAGS < 4 (i.e., allow blends and neighbors, but not saturated detections), (2) IMAFLAGS_ISO = 0 (i.e., remove detections that overlap a CCD defect such as a bad column, a cosmic ray or a stellar bleed trail), and (3) MAG_PSF < 99 (i.e., require the detection has a valid MAG_PSF measurement in a single-epoch image). The NEPOCH variable tracks the number of detections that pass these selection criteria and are used to calculate the WAVG quantities. There are many faint objects that are only detected in the coadded images, and thus have NEPOCH = 0 and no valid WAVG measurements in one or more bands.

We calculate the WAVG_MAG_PSF from the individual MAG_PSF measurements, $i$, in a given band as

$$\text{WAVG\_MAG\_PSF} = \frac{\sum_{i=1}^{i=\text{NEPOCH}}(\text{MAG\_PSF}_i \times w_i)}{\sum_{i=1}^{i=\text{NEPOCH}} w_i}. \quad \text{(A1)}$$

Each measurement is weighted by a factor

$$w_i = ((\text{MAGERR\_PSF})^2 + (0.001)^2)^{-1}. \quad \text{(A2)}$$

The constant term 0.001 is an "error floor" determined empirically to avoid underestimating the errors on very bright objects. The error on the WAVG_MAG_PSF quantity is calculated as

$$\text{WAVG\_MAGERR\_PSF} = \frac{\sqrt{\sum_{i=1}^{i=\text{NEPOCH}}(\text{MAGERR\_PSF}_i)^2}}{\text{NEPOCH}} \quad \text{(A3)}$$

and is a simple measure of the spread in errors for individual detections of a given object in a given band. Similar formulae apply for the calculation of the WAVG_SPREAD_MODEL and WAVG_SPREADERR_MODEL quantities, which are derived from the individual single-epoch SPREAD_MODEL measurements with an error floor of $10^{-6}$. We note that the use of WAVG_MAG_PSF is only recommended for point-like sources.

These definitions of the weighted average (including an error floor) and the error on the average are somewhat nonstandard and have not been robustly verified near the detection limit using fluxes instead of magnitude quantities. However, for stellar objects down to $g \sim 23$, they represent a robust determination of a combined magnitude, which has been used in the past for work on resolved stellar populations and photometric calibration. The main advantage of the WAVG_MAG_PSF quantities for stellar work is that the PSF model is fit on each single-epoch image rather than the coadded image. This makes the WAVG quantities for bright point-like sources more accurate than those derived from the deeper coadded images.

Precise cosmological analyses with DES utilize magnitudes and morphological parameters derived from the simultaneous fitting of the single-epoch images (e.g., the SOF quantities; Drlica-Wagner et al. 2018; Sevilla-Noarbe et al. 2021). These measurements do not suffer from the depth limitations of the WAVG magnitudes, which are derived from independent single-epoch images. We expected that the simultaneously fit quantities derived from the DES DR2 data will be released in support of the DES cosmology analyses in future instances of the DES "Gold" catalog.

## Appendix B
## Photometric Transformations

Here, we provide empirical photometric transformations between DES and several other current and recent sky surveys.

We note that transformation equations can have many forms and are generally dependent on the types of objects for which the photometry is being transformed. Here, we have striven for relatively simple equations based on stars of relatively typical color. As such, these equations should work reasonably well for objects with SEDs not too dissimilar to "normal" stars, but less so for other objects (e.g., objects with strong emission lines). Quality plots and any future updates for these transformations can be found on the DES DR2 web page.[113]

### B.1. SDSS

Here, we matched stars from DES DR2 to stars in the Stripe 82 area of SDSS DR13 (Albareti et al. 2017), using the `WAVG_PSF_MAG`'s from DES and the `PSF_MAG`'s from SDSS. To reduce the effects of photon noise, we only considered stars with rms magnitude errors of $\leqslant 0.01$ in DES and $\leqslant 0.02$ in SDSS. For outlier rejection, we made use of iterated sigma clipping, iterating over the fit $3\times$ and removing $3\sigma$ outliers after each iteration. Finally, we found

$$
\begin{aligned}
g_{\rm DES} &= g_{\rm SDSS} - 0.061(g-i)_{\rm SDSS} + 0.008 \text{ (rms: 0.017 mag)} \\
r_{\rm DES} &= r_{\rm SDSS} - 0.155(r-i)_{\rm SDSS} - 0.007 \text{ (rms: 0.014 mag)} \\
i_{\rm DES} &= i_{\rm SDSS} - 0.166(r-i)_{\rm SDSS} + 0.032 \text{ (rms: 0.018 mag)} \\
z_{\rm DES} &= z_{\rm SDSS} - 0.056(r-i)_{\rm SDSS} + 0.027 \text{ (rms: 0.018 mag)}
\end{aligned}
\tag{B1}
$$

$$
\begin{aligned}
g_{\rm SDSS} &= g_{\rm DES} + 0.060(g-i)_{\rm DES} - 0.005 \text{ (rms: 0.018 mag)} \\
r_{\rm SDSS} &= r_{\rm DES} + 0.150(r-i)_{\rm DES} + 0.014 \text{ (rms: 0.016 mag)} \\
i_{\rm SDSS} &= i_{\rm DES} + 0.167(r-i)_{\rm DES} - 0.027 \text{ (rms: 0.015 mag)} \\
z_{\rm SDSS} &= z_{\rm DES} + 0.054(r-i)_{\rm DES} - 0.024 \text{ (rms: 0.018 mag)},
\end{aligned}
\tag{B2}
$$

i.e., first-order polynomials based on a single color index ($g-i$ for $g$ and $r-i$ for $r$, $i$, $z$). These equations are valid for stars with $-1.0 \lesssim g-i \lesssim 3.5$ ($g$) and $-0.4 \lesssim r-i \lesssim 2.0$ ($r$, $i$, $z$). (Note: the rms listed after each transformation is the rms per star. The mean rms for a collection of stars being transformed from one photometric system to the other—especially if that collection of stars covers a range of colors—should be correspondingly smaller, $\sim\text{rms}/\sqrt{N_{\rm stars}}$.)

### B.2. Pan-STARRS

As for the DES/SDSS transformation equations, here, we matched stars from DES DR2 to stars in the Stripe 82 area of PanSTARRS1 DR2 (Chambers et al. 2016; Flewelling et al. 2020), using the `WAVG_PSF_MAG`'s from DES and the `MEANPSFMAG`'s from PanSTARRS1. Again, to reduce the effects of photon noise, we only considered stars with rms magnitude errors of $\leqslant 0.01$ in DES and $\leqslant 0.02$ in Pan-STARRS1. Using the same type of outlier rejection and aiming for similar goals in accuracy, simplicity, and color

---
[113] https://des.ncsa.illinois.edu/releases/dr2

coverage, we arrived at the following equations:

$$
\begin{aligned}
g_{\rm DES} &= g_{\rm PS1} + 0.028(g-i)_{\rm PS1} + 0.020 \text{ (rms: 0.017 mag)} \\
r_{\rm DES} &= r_{\rm PS1} - 0.142(r-i)_{\rm PS1} - 0.010 \text{ (rms: 0.013 mag)} \\
i_{\rm DES} &= i_{\rm PS1} - 0.155(r-i)_{\rm PS1} + 0.015 \text{ (rms: 0.012 mag)} \\
z_{\rm DES} &= z_{\rm PS1} - 0.114(r-i)_{\rm PS1} - 0.010 \text{ (rms: 0.015 mag)} \\
Y_{\rm DES} &= y_{\rm PS1} - 0.031(r-i)_{\rm PS1} + 0.035 \text{ (rms: 0.017 mag)}
\end{aligned}
\tag{B3}
$$

$$
\begin{aligned}
g_{\rm PS1} &= g_{\rm DES} - 0.026(g-i)_{\rm DES} - 0.020 \text{ (rms: 0.017 mag)} \\
r_{\rm PS1} &= r_{\rm DES} + 0.139(r-i)_{\rm DES} + 0.014 \text{ (rms: 0.015 mag)} \\
i_{\rm PS1} &= i_{\rm DES} + 0.153(r-i)_{\rm DES} - 0.011 \text{ (rms: 0.010 mag)} \\
z_{\rm PS1} &= z_{\rm DES} + 0.112(r-i)_{\rm DES} + 0.013 \text{ (rms: 0.015 mag)} \\
y_{\rm PS1} &= Y_{\rm DES} + 0.031(r-i)_{\rm DES} - 0.034 \text{ (rms: 0.017 mag)}
\end{aligned}
\tag{B4}
$$

which are valid for stars with $-0.9 \lesssim g-i \lesssim 3.8$ ($g$) and $-0.4 \lesssim r-i \lesssim 2.7$ ($r$, $i$, $z$, $Y$).

### B.3. HSC

As described in Section 4.5, we used HSC-PDR2 (Aihara et al. 2019) to report our detection efficiency as a function of DES `MAG_AUTO`. In order to be comparable, transformation equations were derived for `MAG_AUTO` from matched stars between both surveys for high-S/N, high-confidence stars ($18 < g, r, i < 21$) and with $0.2 < g-r < 1.2$ for all bands. We also removed $2\sigma$ outliers after a first fit iteration.

$$
\begin{aligned}
g_{\rm DES} &= g_{\rm HSC} - 0.012(g-r)_{\rm HSC} + 0.029 \text{ (rms: 0.027 mag)} \\
r_{\rm DES} &= r_{\rm HSC} - 0.075(g-r)_{\rm HSC} + 0.031 \text{ (rms: 0.030 mag)} \\
i_{\rm DES} &= i_{\rm HSC} - 0.170(i-z)_{\rm HSC} + 0.022 \text{ (rms: 0.026 mag)} \\
z_{\rm DES} &= z_{\rm HSC} - 0.081(i-z)_{\rm HSC} + 0.023 \text{ (rms: 0.033 mag)} \\
Y_{\rm DES} &= Y_{\rm HSC} + 0.011(i-z)_{\rm HSC} + 0.077 \text{ (rms: 0.057 mag)}
\end{aligned}
\tag{B5}
$$

### B.4. CFHTLenS

In DR1, DES detection efficiency was assessed against CFHTLens data (Erben et al. 2013). For completeness, we also present here the photometric transformations from CFHTLenS to the DES DR2 photometry. Similarly to what was done with HSC-PDR2, transformation equations were derived for `MAG_AUTO` from matched stars between both surveys for high-S/N, high-confidence stars ($18 < g, r, i < 21$) with $0.2 < g-r < 1.2$ and $2\sigma$ outliers removed from the fit:

$$
\begin{aligned}
g_{\rm DES} &= g_{\rm CFHTLenS} + 0.019(g-r)_{\rm CFHTLenS} + 0.117 \text{ (rms: 0.035 mag)} \\
r_{\rm DES} &= r_{\rm CFHTLenS} - 0.062(g-r)_{\rm CFHTLenS} + 0.073 \text{ (rms: 0.025 mag)} \\
i_{\rm DES} &= i_{\rm CFHTLenS} - 0.142(i-z)_{\rm CFHTLenS} + 0.099 \text{ (rms: 0.029 mag)} \\
z_{\rm DES} &= z_{\rm CFHTLenS} + 0.002(i-z)_{\rm CFHTLenS} + 0.099 \text{ (rms: 0.038 mag)}
\end{aligned}
\tag{B6}
$$

### B.5. Johnson–Cousins

Fitting transformation equations between the DES and the Johnson–Cousins ($UBVR_cI_c$) system includes some additional complications. One complication is that the two photometric systems are sufficiently different that the $\Delta$mag versus color plots used for the fits start to have the appearance of full-blown

color–magnitude diagrams—with large discontinuities and occasional bifurcations in the stellar locus. To address this problem, we fit over two or three piecewise discontinuous regions along the color axis and try to avoid band/color combinations that show noticeable bifurcations. The second complication is identifying a large, publicly accessible sample of well-calibrated faint ($V \gtrsim 16$) stellar photometry for the Johnson–Cousins system that overlaps the DES DR2 footprint. We solved this by making use of Peter Stetson's database of Johnson–Cousins standard stars (Stetson 2009; Stetson et al. 2019).[114] We note that Stetson's photometry is calibrated to the photometric system defined by Landolt (1992) and is zero-pointed to the Vega system rather than the *AB* system. Other than these complications, the process was very similar to fitting the other transformation equations in this section. We matched stars from DES DR2 to stars in the Stetson database, using the WAVG_PSF_MAG's from DES and the mean magnitudes from Stetson. To reduce the effects of photon noise, we only considered stars with rms magnitude errors of $\leqslant 0.02$ in either DES or the Stetson sample, and we only included Stetson stars that had at least 5 observations each in *B, V, R, I*. These cuts left us with ≈10,000 matches. As before, for outlier rejection, we made use of iterated sigma clipping, iterating over the fit 3× and removing $3\sigma$ outliers after each iteration. Here are the results

Note that each of the two or three piecewise branches in these relations is discontinuous; no effort was made to ensure that the value obtained from the red endpoint of one branch would match the value of the blue endpoint of the next branch. This was done on purpose, due to the minor discontinuities and/or bifurcations seen in the stellar locus of the plotted relations. We also note that, due to a moderately large bifurcation in the stellar locus, the reddest piecewise branch ($0.7 < (g-r)_{\rm DES} \leqslant 1.8$) of the $B \to g_{\rm DES}$ transformation should be used with some caution (see the quality plots for these relations at the DES DR2 website[115]). Finally, we exclude transformations to/from the Johnson *U* band, because DES DR2 does not contain DECam *u*-band observations.

## Appendix C
## Absolute Calibrations

As with DES DR1, the absolute photometric calibration of DR2 is tied to the HST CALSPEC standard star C26202 (Bohlin et al. 2014), wherein DES *grizY* synthetic *AB* magnitudes were calculated by integrating the DES Standard bandpasses (Figure 5) with the HST CALSPEC c26202_stisnic_007 optical/NIR spectrum of C26202

$$
\begin{aligned}
g_{\rm DES} &= V + 0.552(B-V) - 0.099 \ [-0.2 < (B-V) \leqslant 0.4] \text{(rms: 0.012 mag)} \\
&= V + 0.493(B-V) - 0.067 [0.4 < (B-V) \leqslant 2.2] \text{ (rms: 0.021 mag)} \\
r_{\rm DES} &= R_c + 0.046(R-I)_c + 0.160 \ [-0.2 < (R-I)_c \leqslant 0.7] \text{ (rms: 0.015 mag)} \\
&= R_c + 0.127(R-I)_c + 0.113, \ [0.7 < (R-I)_c \leqslant 2.0] \text{ (rms: 0.021 mag)} \\
i_{\rm DES} &= I_c + 0.083(R-I)_c + 0.392 \ \ [-0.2 < (R-I)_c \leqslant 0.7] \text{ (rms: 0.012 mag)} \\
&= I_c + 0.049(R-I)_c + 0.416 \ \ [0.7 < (R-I)_c \leqslant 2.0] \text{ (rms: 0.015 mag)} \\
z_{\rm DES} &= I_c - 0.546(R-I)_c + 0.494 \ \ [-0.2 < (R-I)_c \leqslant 0.2] \text{(rms: 0.044 mag)} \\
&= I_c - 0.546(R-I)_c + 0.574 \ \ [0.2 < (R-I)_c \leqslant 0.7] \text{ (rms: 0.020 mag)} \\
&= I_c - 0.341(R-I)_c + 0.446 \ \ [0.7 < (R-I)_c \leqslant 2.0] \text{ (rms: 0.023 mag)} \\
Y_{\rm DES} &= I_c - 0.818(R-I)_c + 0.588 \ \ [-0.2 < (R-I)_c \leqslant 0.2] \text{ (rms: 0.051 mag)} \\
&= I_c - 0.831(R-I)_c + 0.698 \ \ [0.2 < (R-I)_c \leqslant 0.7] \text{(rms: 0.026 mag)} \\
&= I_c - 0.437(R-I)_c + 0.451 \ \ [0.7 < (R-I)_c \leqslant 2.0] \text{ (rms: 0.030 mag)}
\end{aligned} \quad \text{(B7)}
$$

$$
\begin{aligned}
B &= g_{\rm DES} + 0.371(g-r)_{\rm DES} + 0.197 \ [-0.5 < (g-r)_{\rm DES} \leqslant 0.2] \text{ (rms: 0.022 mag)} \\
&= g_{\rm DES} + 0.542(g-r)_{\rm DES} + 0.141 \ [0.2 < (g-r)_{\rm DES} \leqslant 0.7] \text{ (rms: 0.017 mag)} \\
&= g_{\rm DES} + 0.454(g-r)_{\rm DES} + 0.200 \ [0.7 < (g-r)_{\rm DES} \leqslant 1.8] \text{ (rms: 0.059 mag)} \\
V &= g_{\rm DES} - 0.465(g-r)_{\rm DES} - 0.020 \ [-0.5 < (g-r)_{\rm DES} \leqslant 0.2] \text{ (rms: 0.012 mag)} \\
&= g_{\rm DES} - 0.496(g-r)_{\rm DES} - 0.015 \ [0.2 < (g-r)_{\rm DES} \leqslant 0.7] \text{ (rms: 0.011 mag)} \\
&= g_{\rm DES} - 0.445(g-r)_{\rm DES} - 0.062 \ [0.7 < (g-r)_{\rm DES} \leqslant 1.8] \text{ (rms: 0.024 mag)} \\
R_c &= r_{\rm DES} - 0.013(r-i)_{\rm DES} - 0.174 \ [-0.4 < (r-i)_{\rm DES} \leqslant 0.1] \text{ (rms: 0.015 mag)} \\
&= r_{\rm DES} - 0.074(r-i)_{\rm DES} - 0.165 \ [0.1 < (r-i)_{\rm DES} \leqslant 0.5] \text{ (rms: 0.013 mag)} \\
&= r_{\rm DES} - 0.120(r-i)_{\rm DES} - 0.149 \ [0.5 < (r-i)_{\rm DES} \leqslant 1.8] \text{ (rms: 0.021 mag)} \\
I_c &= i_{\rm DES} - 0.066(r-i)_{\rm DES} - 0.411 \ [-0.4 < (r-i)_{\rm DES} \leqslant 0.1] \text{ (rms: 0.014 mag)} \\
&= i_{\rm DES} - 0.068(r-i)_{\rm DES} - 0.416 \ [0.1 < (r-i)_{\rm DES} \leqslant 0.5] \text{ (rms: 0.013 mag)} \\
&= i_{\rm DES} - 0.044(r-i)_{\rm DES} - 0.430 \ [0.5 < (r-i)_{\rm DES} \leqslant 1.8] \text{ (rms: 0.016 mag)}
\end{aligned} \quad \text{(B8)}
$$

---

[114] https://www.canfar.net/storage/list/STETSON/Standards

[115] https://des.ncsa.illinois.edu/releases/dr2

Table C1
DES Synthetic AB Magnitudes and Colors for the HST CALSPEC Standard C26202

| Band | `STISNIC.007` | `STISWFCNIC.002` | $\Delta$mag or $\Delta$color |
|---|---|---|---|
| $i$ | 16.2571 | 16.2515 | +0.0056 |
| $g - r$ | 0.3545 | 0.3521 | +0.0024 |
| $r - i$ | 0.0834 | 0.0804 | +0.0030 |
| $i - z$ | 0.0122 | 0.0155 | −0.0033 |
| $z - Y$ | −0.0224 | −0.0133 | −0.0091 |

and converted to magnitudes by applying the standard defining equation for broadband AB magnitudes (Equation (7) of Fukugita et al. 1996). Comparison of the FGCM-calibrated magnitudes of C26202 in each DES band with the corresponding synthetic AB magnitude in that DES band yielded the resulting AB offset in that band, which was then applied to the DES magnitudes.

As noted in Section 4.2.2, there are subtleties involved in that, for the best absolute calibration, one may wish to include an additional AB offset.

One of the subtleties mentioned in Section 4.2.2 is that the HST CALSPEC calibrations for C26202 have changed since DR1: the preferred C26202 spectrum in the current (2020 April 27) HST CALSPEC release is `c26202_stiswfcnic_002`. The DES synthetic AB magnitudes and colors for `c26202_stiswfcnic_002` differ from those of the earlier `c26202_stisnic_007` at the few to several millimagnitude level (see Table C1). If one uses `c26202_stisnic_007` as one's absolute calibration standard, the AB offsets for `WAVG_MAG_PSF` likewise change by the same amount (see Table 2).

In addition, beyond relying on C26202 and a small set of other HST CALSPEC standards that both lie within the DES footprint and do not saturate in the DES science exposures, we have also been building up over the past several years spectroscopic observations of a set of roughly 300 relatively faint (mostly $i \approx 16.5$–18.0) candidate pure-hydrogen-atmosphere ("DA") white dwarfs from various photometric and proper motion catalogs that overlap the DES footprint (Smith et al. 2015). DA white dwarfs are useful spectrophotometric calibrators because their spectra are more or less blackbodies imprinted with broad, deep hydrogen Balmer absorption, making their spectra comparatively easy to model accurately and robustly. It is typically their modeled spectra—not their observed spectra—that are used as the spectrophotometric standards (Holberg 2007; Betoule et al. 2013; Axelrod & Miller 2014; Camarota & Holberg 2014; Narayan et al. 2016; Calamida et al. 2019; López-Sanjuan et al. 2019; Narayan et al. 2019; Wall et al. 2019;

Gentile Fusillo et al. 2020). For our DA sample, modeling was performed using the DA white dwarf models described in Tremblay et al. (2011, 2013). After removing non-DA's, poorly measured DA's, variable stars, etc., we end up with a "Golden Sample" of $\approx 150$ DA white dwarfs with well-modeled spectra. These models provide an excellent measure of the detailed shapes of the DA spectra (Figure 5) from which we obtain accurate and precise synthetic DES AB colors. That said, the models do not contain distance information, so we currently do not have sufficiently accurate and independent measurements of the overall normalization of the individual DA spectra to provide useful AB $i$-band offsets based on the DA sample (Table 2).[116]

As described in Section 4.2.2, the DES DR2 magnitudes and colors as provided are already quite close to the AB magnitude system (offset by $\lesssim 0.03$ mag) and that offsets for `WAVG_MAG_PSF` and for `MAG_AUTO` are very similar (typically within a few millimagnitude) (see Table 2)[117]. That said, we do find a tension between the estimated AB color offsets based on C26202 and those based on the DA sample, particularly for the redder color indices ($i - z$ and $z - Y$).

For those wishing to apply the additional offsets from Table 2 in order to improve the DES DR2 data's tie to the AB system, we note that there are pros and cons associated with using either the C26020 AB calibration or the DA sample AB calibration. The benefit of using the C26020 AB calibration is that it is tied directly to HST observations of a star of average color (F8 IV; roughly solar analog); the benefit of using the DA sample AB calibration is that it is based on the well-modeled spectra of many stars spread over the full DES footprint. In many ways, the two AB calibrations of the DES passbands are complementary, due in part to the different range of colors of the stars in question (see Figure 5). The DES Collaboration does not officially endorse a preference for either of these updated AB offsets at this time but recommends the user to check the DES DR2 website for updates (as well as for offsets for other DES DR2 magnitude types beyond `MAG_AUTO` and `WAVG_MAG_PSF`).

Further details of our CALSPEC- and DA-based calibration programs will be presented in a forthcoming paper.

# Appendix D
# Source Extractor Parameters & Flags

Table D1 summarizes the standard warning flags provided by `SourceExtractor`, encoded in the `FLAGS` bitmask column for each band. Table D2 contains the diameters used for `SourceExtractor` aperture magnitude determination.

---

[116] For those who wish to use the AB offsets from the DA sample, we suggest using the `c26202_stiswfcnic_002` value for the $i$-band AB offset (see Table 2).
[117] The sizes and differences between the `WAVG_MAG_PSF` and `MAG_AUTO` offsets for `c26202_stisnic_007` are indicative of how much different choices for the choice of single-epoch exposures and choice of magnitude type affect the estimates of the AB offsets—i.e., roughly on the order of 1–6 mmag.

**Table D1**
Summary of Bitmask Values and Warning Descriptions for the `SourceExtractor` FLAGS Column[a]

| Bit | Description |
| --- | --- |
| 1 | Aperture photometry is likely to be biased by neighboring sources or by more than 10% of bad pixels in any aperture |
| 2 | The object was originally blended with another one |
| 4 | At least one pixel of the object is saturated (or very close to) |
| 8 | The isophotal footprint of the detected object is truncated (too close to an image boundary) |
| 16 | At least one photometric aperture is incomplete or corrupted (hitting buffer or memory limits) |
| 32 | The isophotal footprint is incomplete or corrupted (hitting buffer or memory limits) |
| 64 | A memory overflow occurred during deblending |
| 128 | A memory overflow occurred during extraction |

**Note.**
[a] Table data obtained from https://sextractor.readthedocs.io/en/latest/Flagging.html.

**Table D2**
Diameters for the Set of Aperture Magnitudes

| Column Name | Diameter (pixels) | Diameter (arcseconds) |
| --- | --- | --- |
| MAG_APER[1] | 1.85 | 0.49 |
| MAG_APER[2] | 3.70 | 0.97 |
| MAG_APER[3] | 5.55 | 1.46 |
| MAG_APER[4] | 7.41 | 1.95 |
| MAG_APER[5] | 11.11 | 2.92 |
| MAG_APER[6] | 14.81 | 3.90 |
| MAG_APER[7] | 18.52 | 4.87 |
| MAG_APER[8] | 22.22 | 5.84 |
| MAG_APER[9] | 25.93 | 6.82 |
| MAG_APER[10] | 29.63 | 7.79 |
| MAG_APER[11] | 44.44 | 11.69 |
| MAG_APER[12] | 66.67 | 17.53 |

## Appendix E
## Released Tables

The DES DR2 catalog data comprise five tables. DR2_MAIN (see Table E1) includes all the main quantities extracted from the coadd pipeline and important information about the objects. That table also includes MAG_AUTO and WAVG_MAG_PSF, associated uncertainties, as well as the corresponding dereddened magnitudes along with star/galaxy separation columns. DR2_FLUX (see Table E2) and DR2_MAGNITUDE (see Table E3) contain 15 different measurements of the fluxes and magnitudes for each object. Additionally, these three tables share some commonly used columns to facilitate queries (by avoiding the need to join multiple tables) like ID, position, HEALPix index, and filename. The fourth table, DR2_COVERAGE (see Table E4), contains information about the survey footprint in HEALPix format. Finally, DR2_TILE_INFO (see Table E5), contains information relevant to the processed tiles, including the tile geometry and the names and URL of the associated files for the images and catalogs.

**Table E1**
DR2_MAIN (Table Description: 691,483,608; 215 Columns)

| Column Name | Description | Columns |
|---|---|---|
| COADD_OBJECT_ID | Unique identifier for the coadded objects | 1 |
| TILENAME | Identifier of each one of the tiles on which the survey is gridded | 1 |
| R.A. | Right ascension, with quantized precision for indexing (ALPHAWIN_J2000 has full precision but not indexed) [deg] | 1 |
| ALPHAWIN_J2000 | Right ascension for the object, J2000 in ICRS system (full precision but not indexed) [deg] | 1 |
| Decl. | decl., with quantized precision for indexing (DELTAWIN_J2000 has full precision but not indexed) [deg] | 1 |
| DELTAWIN_J2000 | decl. for the object, J2000 in ICRS system (full precision but not indexed) [deg] | 1 |
| GALACTIC_L | Galactic Longitude [deg] | 1 |
| GALACTIC_B | Galactic Latitude [deg] | 1 |
| XWIN_IMAGE | X-centroid from windowed measurements on coadded image [pixel] | 1 |
| YWIN_IMAGE | Y-centroid from windowed measurements on coadded image [pixel] | 1 |
| XWIN_IMAGE_G,R,I,Z,Y | X-centroid from windowed measurements on coadded band image [pixel] | 5 |
| YWIN_IMAGE_G,R,I,Z,Y | Y-centroid from windowed measurements on coadded band images [pixel] | 5 |
| X2WIN_IMAGE_G,R,I,Z,Y | Second moment in $x$-direction, from converged windowed measurements [pixel2] | 5 |
| ERRX2WIN_IMAGE_G,R,I,Z,Y | Uncertainty in second moment of $x$-distribution centroid, from converged windowed measurements [pixel2] | 5 |
| Y2WIN_IMAGE_G,R,I,Z,Y | Second moment in $y$-direction, from converged windowed measurements [pixel2] | 5 |
| ERRY2WIN_IMAGE_G,R,I,Z,Y | Uncertainty in second moment of $y$-distribution centroid, from converged windowed measurements [pixel2] | 5 |
| XYWIN_IMAGE_G,R,I,Z,Y | Second moment in $xy$-direction, from converged windowed measurements [pixel2] | 5 |
| ERRXYWIN_IMAGE_G,R,I,Z,Y | Uncertainty in second moment of $xy$-distribution, from converged windowed measurements [pixel2] | 5 |
| HPIX_32,64,1024,4096,16384 | Healpix identifier for its `nside` grid size, in a NESTED schema | 5 |
| NEPOCHS_G,R,I,Z,Y | Number of epochs the source is detected in single-epoch images | 5 |
| NITER_MODEL_G,R,I,Z,Y | Number of iterations in model fitting photometric measurements | 5 |
| ISOAREA_IMAGE_G,R,I,Z,Y | Isophotal area of the coadded source [pixel2] | 5 |
| A_IMAGE | Major axis size based on an isophotal model [pixel] | 1 |
| ERRA_IMAGE | Uncertainty in major axis size, from isophotal model [pixel] | 1 |
| AWIN_IMAGE_G,R,I,Z,Y | Major axis size, from 2nd order windowed moment measurements [pixels] | 5 |
| ERRAWIN_IMAGE_G,R,I,Z,Y | Uncertainty in major axis size, from converged windowed measurement, assuming uncorrelated noise [pixel] | 5 |
| B_IMAGE | Minor axis size based on an isophotal model [pixel] | 1 |
| ERRB_IMAGE | Uncertainty in minor axis size, from isophotal model [pixel] | 1 |
| BWIN_IMAGE_G,R,I,Z,Y | Minor axis size, from 2nd order windowed moment measurements [pixels] | 5 |
| ERRBWIN_IMAGE_G,R,I,Z,Y | Uncertainty in minor axis size, from converged windowed measurement, assuming uncorrelated noise [pixel] | 5 |
| THETA_J2000 | Position angle of source in J2000 coordinates, from nonwindowed measurement [deg] | 1 |
| ERRTHETA_IMAGE | Uncertainty in source position, from isophotal model [deg] | 1 |
| THETAWIN_IMAGE_G,R,I,Z,Y | Position angle of source, for converged windowed measurement grow from $x$ to $y$ [deg] | 5 |
| ERRTHETAWIN_IMAGE_G,R,I,Z,Y | Uncertainty in source position angle, from converged windowed measurement [deg] | 5 |
| FWHM_IMAGE_G,R,I,Z,Y | FWHM measured from the isophotal area, from elliptical growth-curve, modeled in two-dimensions [pixel] | 5 |
| FLUX_RADIUS_G,R,I,Z,Y | Half-light radius for the object, from elliptical growth-curve, modeled in two-dimensions [pixel] | 5 |
| KRON_RADIUS | Kron radius measured from detection image [pixel] | 1 |
| KRON_RADIUS_G,R,I,Z,Y | Kron radius measured from coadded image [pixel] | 5 |
| CLASS_STAR_G,R,I,Z,Y | Simple morphological extended source classifier. Values between 0 (galaxies) and 1 (stars). SPREAD_MODEL exhibits better performance for morphological classification. | 5 |
| SPREAD_MODEL_G,R,I,Z,Y | Morphology based classifier based on comparison between a PSF versus exponential-PSF model. Values closer to 0 correspond to stars, larger values correspond to galaxies | 5 |
| SPREADERR_MODEL_G,R,I,Z,Y | Uncertainty in morphology based classifier based on comparison between PSF versus exponential-PSF model. | 5 |
| WAVG_SPREAD_MODEL_G,R,I,Z,Y | SPREAD MODEL using the weighted averaged values from single-epoch detections | 5 |
| WAVG_SPREADERR_MODEL_G,R,I,Z,Y | Uncertainty in SPREAD MODEL using the weighted averaged values from single-epoch detections | 5 |
| FLUX_AUTO_G,R,I,Z,Y | Aperture-flux measurement, elliptical model based on the Kron radius [ADU] | 5 |
| FLUXERR_AUTO_G,R,I,Z,Y | Uncertainty in aperture-flux measurement, elliptical model based on the Kron radius [ADU] | 5 |
| WAVG_FLUX_PSF_G,R,I,Z,Y | Weighted-average flux measurement of PSF fit single-epoch detections [ADU] | 5 |
| WAVG_FLUXERR_PSF_G,R,I,Z,Y | Uncertainty of weighted averaged flux measurement of PSF fit single-epoch detections [ADU] | 5 |
| MAG_AUTO_G,R,I,Z,Y | Magnitude estimation, for an elliptical model based on the Kron radius [mag] | 5 |
| MAGERR_AUTO_G,R,I,Z,Y | Uncertainty in magnitude estimation, for an elliptical model based on the Kron radius [mag] | 5 |
| MAG_AUTO_G,R,I,Z,Y_DERED | Dereddened magnitude estimation (using SFD98), for an elliptical model based on the Kron radius [mag] | 5 |
| WAVG_MAG_PSF_G,R,I,Z,Y | Weighted-average magnitude, of PSF fit single-epoch detections [mag] | 5 |

**Table E1**
(Continued)

| Column Name | Description | Columns |
|---|---|---|
| WAVG_MAGERR_PSF_G,R,I,Z,Y | Uncertainty of weighted averaged magnitude, of PSF fit single-epoch detections [mag] | 5 |
| WAVG_MAG_PSF_G,R,I,Z,Y_DERED | Dereddened weighted-average magnitude (using SFD98) from PSF fit single-epoch detections [mag] | 5 |
| EBV_SFD98 | E(B-V) reddening coefficient from Schlegel, Finkbeiner & Davis, 1998 [mag] | 1 |
| BACKGROUND_G,R,I,Z,Y | Background level, by CCD-amplifier [mag] | 5 |
| FLAGS_G,R,I,Z,Y | Flag describing cautionary advice about source extraction process (FLAGS < 4 for well-behaved objects) | 5 |
| IMAFLAGS_ISO_G,R,I,Z,Y | Flag identifying sources with missing/flagged pixels, considering all single-epoch images | 5 |
| EXTENDED_CLASS_COADD | 0: high-confidence stars; 1: candidate stars; 2: mostly gxs; 3: high-confidence gxs; -9: No data; Sextractor photometry | 1 |
| EXTENDED_CLASS_WAVG | 0: high-confidence stars; 1: candidate stars; 2: mostly gxs; 3: high-confidence gxs; -9: No data; WAVG photometry | 1 |

**Table E2**
DR2_FLUX (Table Description: 691,483,608, 179 Columns)

| Column Name | Description | Columns |
|---|---|---|
| COADD_OBJECT_ID | Unique identifier for the coadded objects | 1 |
| TILENAME | Identifier of each one of the tiles on which the survey is gridded | 1 |
| R.A. | Right ascension, with quantized precision for indexing (ALPHAWIN_J2000 has full precision but not indexed) [deg] | 1 |
| ALPHAWIN_J2000 | Right ascension for the object, J2000 in ICRS system (full precision but not indexed) [deg] | 1 |
| Decl. | decl., with quantized precision for indexing (DELTAWIN_J2000 has full precision but not indexed) [deg] | 1 |
| DELTAWIN_J2000 | decl. for the object, J2000 in ICRS system (full precision but not indexed) [deg] | 1 |
| XWIN_IMAGE | X-centroid of source (from coadd detection image) [pix] | 1 |
| YWIN_IMAGE | Y-centroid of source (from coadd detection image) [pix] | 1 |
| HPIX_32,64,1024,4096,16384 | Healpix identifier for its nside grid size, in a NESTED schema | 5 |
| FLUX_AUTO_G,R,I,Z,Y | Flux measurement, for an elliptical model based on the Kron radius [ADU] | 5 |
| FLUXERR_AUTO_G,R,I,Z,Y | Uncertainty in flux measurement, for an elliptical model based on the Kron radius [ADU] | 5 |
| FLUX_APER_1-12_G,R,I,Z,Y | Flux measurement for circular apertures [ADU] | 60 |
| FLUXERR_APER_1-12_G,R,I,Z,Y | Uncertainty in flux measurement for circular apertures [ADU] | 60 |
| FLUX_PETRO_G,R,I,Z,Y | Flux for a Petrosian radius [ADU] | 5 |
| FLUXERR_PETRO_G,R,I,Z,Y | Uncertainty in flux for a Petrosian radius [ADU] | 5 |
| WAVG_FLUX_PSF_G,R,I,Z,Y | Weighted averaged flux, of PSF fit single-epoch detections [ADU] | 5 |
| WAVG_FLUXERR_PSF_G,R,I,Z,Y | Uncertainty of weighted averaged flux, of PSF fit single-epoch detections [ADU] | 5 |
| PETRO_RADIUS_G,R,I,Z,Y | Petrosian radius [pix] | 5 |
| EBV_SFD98 | $E(B-V)$ reddening coefficient from Schlegel, Finkbeiner & Davis, 1998 [mag] | 1 |
| FLAGS_G,R,I,Z,Y | Additive flag describing cautionary advice about source extraction process. Use less than 4 for well-behaved objects | 5 |
| IMAFLAGS_ISO_G,R,I,Z,Y | Flag identifying sources with missing/flagged pixels, considering all single-epoch images | 5 |

**Table E3**
DR2_MAGNITUDE (Table Description: 691,483,608 Rows, 179 Columns)

| Column Name | Description | Columns |
|---|---|---|
| COADD_OBJECT_ID | Unique identifier for the coadded objects | 1 |
| TILENAME | Identifier of each one of the tiles on which the survey is gridded | 1 |
| R.A. | Right ascension, with quantized precision for indexing (ALPHAWIN_J2000 has full precision but not indexed) [deg] | 1 |
| ALPHAWIN_J2000 | Right ascension for the object, J2000 in ICRS system (full precision but not indexed) [deg] | 1 |
| Decl. | decl., with quantized precision for indexing (DELTAWIN_J2000 has full precision but not indexed) [deg] | 1 |
| DELTAWIN_J2000 | decl. for the object, J2000 in ICRS system (full precision but not indexed) [deg] | 1 |
| XWIN_IMAGE | X-centroid from windowed measurements on coadded image [pixel] | 1 |
| YWIN_IMAGE | Y-centroid from windowed measurements on coadded image [pixel] | 1 |
| HPIX_32,64,1024,4096,16384 | Healpix identifier for its nside grid size, in a NESTED schema | 5 |
| MAG_AUTO_G,R,I,Z,Y | Magnitude estimation, for an elliptical model based on the Kron radius [mag] | 5 |
| MAGERR_AUTO_G,R,I,Z,Y | Uncertainty in magnitude estimation, for an elliptical model based on the Kron radius [mag] | 5 |
| MAG_APER_1-12_G,R,I,Z,Y | Magnitude estimation for circular apertures [mag] | 60 |
| MAGERR_APER_1-12_G,R,I,Z,Y | Uncertainty in magnitude estimation for circular apertures [mag] | 60 |
| MAG_PETRO_G,R,I,Z,Y | Magnitude for a Petrosian radius [mag] | 5 |
| MAGERR_PETRO_G,R,I,Z,Y | Uncertainty in magnitude for a Petrosian radius [mag] | 5 |

Table E3
(Continued)

| Column Name | Description | Columns |
|---|---|---|
| WAVG_MAG_PSF_G,R,I,Z,Y | Weighted-average magnitude, of PSF fit single-epoch detections [mag] | 5 |
| WAVG_MAGERR_PSF_G,R,I,Z,Y | Uncertainty of weighted averaged magnitude, of PSF fit single-epoch detections [mag] | 5 |
| PETRO_RADIUS_G,R,I,Z,Y | Petrosian radius [pixel] | 5 |
| EBV_SFD98 | $E(B-V)$ reddening coefficient from Schlegel et al. (1998) [mag] | 1 |
| FLAGS_G,R,I,Z,Y | Additive flag describing cautionary advice about source extraction process. Use fewer than four for well-behaved objects | 5 |
| IMAFLAGS_ISO_G,R,I,Z,Y | Flag identifying sources with missing/flagged pixels, considering all single-epoch images | 5 |

Table E4
DR2_COVERAGE (Table Description: 25,239,595 Rows, 6 Columns)

| Column Name | Description | Columns |
|---|---|---|
| HPIX_4096 | Unique `HEALPix` pixel ID, in NESTED ordering at `nside = 4096` | 1 |
| FRAC_DET_G,R,I,Z,Y | Fractional area of Healpix pixel covered by the individual band | 5 |

Table E5
DR2_TILE_INFO (Table Description: 10,169 Rows, 53 Columns)

| Column Name | Description | Columns |
|---|---|---|
| TILENAME | Tile-name identifier | 1 |
| ID | Internal unique ID identifier | 1 |
| RA_CENT | Central R.A. for tile [deg] | 1 |
| DEC_CENT | Central decl. for tile [deg] | 1 |
| COUNT | Number of objects per tile | 1 |
| RAC1 | R.A. at Corner 1 of tile [deg] | 1 |
| RAC2 | R.A. at Corner 2 of tile [deg] | 1 |
| RAC3 | R.A. at Corner 3 of tile [deg] | 1 |
| RAC4 | R.A. at Corner 4 of tile [deg] | 1 |
| RACMAX | Maximum R.A. covered in tile [deg] | 1 |
| RACMIN | Minimum R.A. covered in tile [deg] | 1 |
| RA_SIZE | Extent of R.A. for tile [deg] | 1 |
| URAMAX | Maximum Unique R.A. of objects measured from tile [deg] | 1 |
| URAMIN | Minimum Unique R.A. of objects measured from tile [deg] | 1 |
| DECC1 | decl. at Corner 1 of tile [deg] | 1 |
| DECC2 | decl. at Corner 2 of tile [deg] | 1 |
| DECC3 | decl. at Corner 3 of tile [deg] | 1 |
| DECC4 | decl. at Corner 4 of tile [deg] | 1 |
| DECCMAX | Maximum decl. covered in tile [deg] | 1 |
| DECCMIN | Minimum decl. covered in tile [deg] | 1 |
| DEC_SIZE | Extent of decl. for tile, in average is 0.7304 deg [deg] | 1 |
| UDECMAX | Maximum Unique decl. of objects measured from tile [deg] | 1 |
| UDECMIN | Minimum Unique decl. of objects measured from tile [deg] | 1 |
| CTYPE1 | WCS projection used for axis 1. Value: RA—TAN | 1 |
| CTYPE2 | WCS projection used for axis 2. Value: DEC–TAN | 1 |
| NAXIS1 | WCS definition for number of pixels for axis 1 | 1 |
| NAXIS2 | WCS definition for number of pixels for axis 2 | 1 |
| CRPIX1 | WCS definition of central pixel for axis 1. Value: 5000.5 | 1 |
| CRPIX2 | WCS definition of central pixel for axis 2. Value: 5000.5 | 1 |
| CRVAL1 | WCS definition of central pixel value for axis 1 | 1 |
| CRVAL2 | WCS definition of central pixel value for axis 2 | 1 |
| CD1_1 | WCS definition for pixel orientation. Value: -0.0000730556 | 1 |
| CD1_2 | WCS definition for pixel orientation. Value: 0 | 1 |
| CD2_1 | WCS definition for pixel orientation. Value: 0 | 1 |
| CD2_2 | WCS definition for pixel orientation. Value: 0.0000730556 | 1 |
| CROSSRA0 | Flag tile boundary crosses RA=0/24h boundary. Values: Y or N | 1 |
| PIXELSCALE | WCS definition of pixel scale. Values: 0.263 arcsec/pix [arcsec/pix] | 1 |

Table E5
(Continued)

| Column Name | Description | Columns |
|---|---|---|
| TIFF_COLOR_IMAGE | Filename of the TIFF image for the tile, being tilename_r{reqnum}p{attnum}_irg.tiff | 1 |
| TIFF_COLOR_IMAGE_NOBKG | Filename of the TIFF image for the tile (No background removed), being tilename_r{reqnum}p{attnum}_irg_nobkg.tiff | 1 |
| FITS_DR2_FLUX | Filename of the served FITS being tilename_dr2_flux.fits | 1 |
| FITS_DR2_MAGNITUDE | Filename of the served FITS being tilename_dr2_magnitude.fits | 1 |
| FITS_DR2_MAIN | Filename of the served FITS being tilename_dr2_main.fits | 1 |
| FITS_IMAGE_DET | Filename of the served FITS being tilename_r{reqnum}p{attnum}_det.fits.fz | 1 |
| FITS_IMAGE_G,R,I,Z,Y | Filename of the served FITS, per band, being tilename_r{reqnum}p{attnum}_band.fits.fz | 5 |
| FITS_IMAGE_NOBKG_G,R,I,Z,Y | Filename of the served FITS, No background removed, per band, being tilename_r{reqnum}p{attnum}_band_nobkg.fits.fz | 5 |

**ORCID iDs**

M. Adamów https://orcid.org/0000-0002-6904-359X
M. Aguena https://orcid.org/0000-0001-5679-6747
S. Allam https://orcid.org/0000-0002-7069-7857
J. Annis https://orcid.org/0000-0002-0609-3987
M. R. Becker https://orcid.org/0000-0001-7774-2246
G. M. Bernstein https://orcid.org/0000-0002-8613-8259
E. Bertin https://orcid.org/0000-0002-3602-3664
S. L. Bridle https://orcid.org/0000-0002-0128-1006
D. Brooks https://orcid.org/0000-0002-8458-5047
D. L. Burke https://orcid.org/0000-0003-1866-1950
A. Carnero Rosell https://orcid.org/0000-0003-3044-5150
M. Carrasco Kind https://orcid.org/0000-0002-4802-3194
J. Carretero https://orcid.org/0000-0002-3130-0204
R. Cawthon https://orcid.org/0000-0003-2965-6786
C. Chang https://orcid.org/0000-0002-7887-0896
A. Choi https://orcid.org/0000-0002-5636-233X
C. Conselice https://orcid.org/0000-0003-1949-7638
M. Crocce https://orcid.org/0000-0002-9745-6228
L. N. da Costa https://orcid.org/0000-0002-7731-277X
T. M. Davis https://orcid.org/0000-0002-4213-8783
J. De Vicente https://orcid.org/0000-0001-8318-6813
S. Desai https://orcid.org/0000-0002-0466-3288
H. T. Diehl https://orcid.org/0000-0002-8357-7467
A. Drlica-Wagner https://orcid.org/0000-0001-8251-933X
K. Eckert https://orcid.org/0000-0002-1407-4700
J. Elvin-Poole https://orcid.org/0000-0001-5148-9203
A. E. Evrard https://orcid.org/0000-0002-4876-956X
A. Ferté https://orcid.org/0000-0003-3065-9941
P. Fosalba https://orcid.org/0000-0002-1510-5214
D. Friedel https://orcid.org/0000-0002-3632-7668
J. Frieman https://orcid.org/0000-0003-4079-3263
J. García-Bellido https://orcid.org/0000-0002-9370-8360
E. Gaztanaga https://orcid.org/0000-0001-9632-0815
D. W. Gerdes https://orcid.org/0000-0001-6942-2736
T. Giannantonio https://orcid.org/0000-0002-9865-0436
M. S. S. Gill https://orcid.org/0000-0003-2524-5154
D. Gruen https://orcid.org/0000-0003-3270-7644
R. A. Gruendl https://orcid.org/0000-0002-4588-6517
J. Gschwend https://orcid.org/0000-0003-3023-8362
G. Gutierrez https://orcid.org/0000-0003-0825-0517
S. R. Hinton https://orcid.org/0000-0003-2071-9349
D. L. Hollowood https://orcid.org/0000-0002-9369-4157
K. Honscheid https://orcid.org/0000-0002-6550-2023
D. Huterer https://orcid.org/0000-0001-6558-0112
D. J. James https://orcid.org/0000-0001-5160-4486
M. D. Johnson https://orcid.org/0000-0002-8290-0533
S. Kent https://orcid.org/0000-0003-4207-7420
K. Kuehn https://orcid.org/0000-0003-2511-0946
N. Kuropatkin https://orcid.org/0000-0003-2511-0946
O. Lahav https://orcid.org/0000-0002-1134-9035
T. S. Li https://orcid.org/0000-0002-9110-6163
C. Lidman https://orcid.org/0000-0003-1731-0497
H. Lin https://orcid.org/0000-0002-7825-3206
N. MacCrann https://orcid.org/0000-0002-8998-3909
M. A. G. Maia https://orcid.org/0000-0001-9856-9307
T. A. Manning https://orcid.org/0000-0003-2545-9195
J. L. Marshall https://orcid.org/0000-0003-0710-9474
P. Martini https://orcid.org/0000-0002-4279-4182
P. Melchior https://orcid.org/0000-0002-8873-5065
F. Menanteau https://orcid.org/0000-0002-1372-2534
R. Miquel https://orcid.org/0000-0002-6610-4836
R. Morgan https://orcid.org/0000-0002-7016-5471
J. Myles https://orcid.org/0000-0001-6145-5859
E. Neilsen https://orcid.org/0000-0002-7357-0317
R. L. C. Ogando https://orcid.org/0000-0003-2120-1154
A. Palmese https://orcid.org/0000-0002-6011-0530
A. Pieres https://orcid.org/0000-0001-9186-6042
A. A. Plazas https://orcid.org/0000-0002-2598-0514
A. K. Romer https://orcid.org/0000-0002-9328-879X
A. Roodman https://orcid.org/0000-0001-5326-3486
E. S. Rykoff https://orcid.org/0000-0001-9376-3135
M. Sako https://orcid.org/0000-0003-2764-7093
E. Sanchez https://orcid.org/0000-0002-9646-8198
J. Allyn Smith https://orcid.org/0000-0002-6261-4601
M. Smith https://orcid.org/0000-0002-3321-1432
M. E. C. Swanson https://orcid.org/0000-0002-1488-8552
G. Tarle https://orcid.org/0000-0003-1704-0781
D. Thomas https://orcid.org/0000-0002-6325-5671
D. L. Tucker https://orcid.org/0000-0001-7211-5729
A. R. Walker https://orcid.org/0000-0002-7123-8943
R. H. Wechsler https://orcid.org/0000-0003-2229-011X
J. Weller https://orcid.org/0000-0002-8282-2010
W. Wester https://orcid.org/0000-0003-0072-6736
B. Yanny https://orcid.org/0000-0002-9541-2678
Y. Zhang https://orcid.org/0000-0001-5969-4631
R. Nikutta https://orcid.org/0000-0002-7052-6900
M. Fitzpatrick https://orcid.org/0000-0002-9080-0751
A. Jacques https://orcid.org/0000-0001-9631-831X
A. Scott https://orcid.org/0000-0002-1140-5463
K. Olsen https://orcid.org/0000-0002-7134-8296
L. Huang https://orcid.org/0000-0003-0952-5789

D. Herrera ● https://orcid.org/0000-0003-2092-6727
S. Juneau ● https://orcid.org/0000-0002-0000-2394
D. Nidever ● https://orcid.org/0000-0002-1793-3689